\ifdefined\pdfminorversion\pdfminorversion=6\fi
\documentclass[
  letterpaper,
  aps,
  prl,
  longbibliography,
  reprint,
  twocolumn,
  superscriptaddress,
  floatfix
]{revtex4-2}

\usepackage[T1]{fontenc}
\usepackage{amsmath,amssymb,bm}
\usepackage{mathrsfs}
\usepackage{graphicx}
\usepackage{xcolor}
\usepackage{booktabs,array}
\usepackage{utfsym}
\usepackage[
  colorlinks=true,
  linkcolor=blue,
  citecolor=blue,
  urlcolor=blue,
  pdfstartview=FitV
]{hyperref}

\newcommand{\bk}{\boldsymbol{k}}
\newcommand{\bE}{\boldsymbol{E}}
\newcommand{\Uop}{\mathcal U}
\newcommand{\Dop}{\mathcal D}

\begin{document}

\title{General Transport Signatures of Odd-Parity Magnets: Charge--Spin Selection Rules and Temporal--Spectral Separation}
\author{Junxi Yu}
\author{Pu Zhang}
\author{Yichen Liu}
\author{Yongpan Li}
\author{Cheng-Cheng Liu}
\email{ccliu@bit.edu.cn}
\affiliation{Centre for Quantum Physics, Key Laboratory of Advanced Optoelectronic Quantum Architecture and Measurement (MOE), School of Physics, Beijing Institute of Technology, Beijing 100081, China}

\begin{abstract}
While even-parity altermagnets are known for spatially separating charge and spin currents via the crystal spin Hall effect, the defining transport counterpart for odd-parity magnets remains to be established. Here, we establish temporal and spectral charge--spin separation as a general transport signature across all spin-splitting classes of collinear and noncollinear odd-parity magnets with unitary spin-flip inversion $\mathcal U=[C_2^\perp\Vert P]$. This symmetry restricts charge responses to odd orders and $z$-polarized spin responses to even orders, independent of microscopic mechanism and without requiring spin conservation. For ac drives that reverse sign after half a period, charge and spin currents exhibit opposite half-period parities, and their spectra contain only odd charge and even spin harmonics. This allows dc spin rectification without a dc charge counterpart, even in longitudinal transport and beyond weak fields. We show that although antiunitary $\mathcal T\tau$ enforces odd-parity bands, it reverses the charge--spin selection of Berry-curvature and Berry-curvature-dipole responses, providing a direct transport fingerprint distinguishing the two protecting symmetries. We verify the selection rules for all linear and second-order transport mechanisms and demonstrate harmonic separation in a minimal odd-parity model, representative Floquet-engineered collinear MnPS$_3$, and the equilibrium noncollinear VCl$_2$ family. 
\end{abstract}

\maketitle

\textit{Introduction---}Charge and spin currents are two fundamental transport channels. Efficient electrical generation of spin currents is a central objective in spintronics~\cite{sinovaSpinHallEffects2015,hayamiNonlinearSpinHall2022,FeiSpinCPGE2021,shao_spinneutral_2021}. Even-parity altermagnets provide a nonrelativistic route to spin-current generation through the crystal spin Hall effect, in which crystal symmetry directs opposite spins along different directions~\cite{smejkalBeyondConventional2022,smejkalEmergingResearchLandscape2022,hayamiMomentumDependent2019,YoshidaSpinCPGE2026,FangAltermagnetNonlinear2024,linCoulombDragAltermagnets2025,ChuThirdOrderRuO22025,LyAltermagnetPumping2026,GunninkAcousticSpinSplitter2026,ZhangAnisotropicMagnetoresistance2025,usachevNonlinearOctupolar2026,maHyperspinAltermagnets2026,yangNonlinearOptomagnetic2026,sankarBroadbandNonlinearHall2025,wan_helical_2025,mizuno_highharmonic_2026a,gonzalezHernandezElectricalSpinSplitter2021,laiFlatFermiSurface2025,liu_twisted_2024}. In contrast, odd-parity magnets (OPMs), encompassing both collinear and noncollinear magnetic orders, exhibit opposite spin polarizations at inversion-related momenta~\cite{hayamiAntisymmetricSpinSplitting2020,hellenesPWaveMagnets2023,yuOddParityMagnetism2025,songElectricalSwitching2025,yamadaMetallicPWave2025,chengEvidencePWaveEuSi2026,SimQuantumPWave2026,linSublatticeCurrents2026,leebCollinearPWave2026,zhuangOrbitalOrders2025,zengOddParitySpinGroup2026,luoSpinGroupCriteria2026,huangLightInducedOddParity2026,zhuFloquetOddParity2026,liuLightInducedDimerized2026,zhangTriangulene2026,li_floquet_2026,liu_complete_2026,bai_timereversalodd_2026}. While even-parity altermagnets can separate charge and spin currents in space through the crystal spin Hall effect, a general principle governing their separation in OPMs remains to be established.

Recent studies have revealed nonrelativistic Edelstein effects, shift photocurrents, and higher-order spin currents in OPMs~\cite{chakrabortyNonrelativisticEdelstein2025,brekkeMinimalModels2024,sivianesOpticalSignatures2025,ezawaHigherOrderSpinCurrent2025,luoAnomalousHallOPM2026}. However, these investigations have primarily focused on specific response orders, microscopic mechanisms, spin-splitting classes, or materials. A fundamental question is whether odd-parity symmetry alone can dictate a general, mechanism-independent principle governing charge and spin transport across all response orders and all spin-splitting classes ($l=1,3,\cdots,9$)~\cite{liu_complete_2026,elcoro_spin_2026,songUnifiedMagneticOrders2026}.

\begin{figure}
  \centering
  \includegraphics[width=\columnwidth]{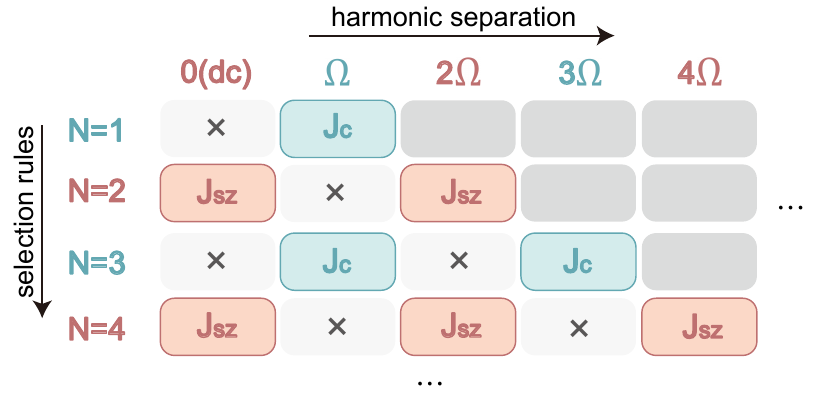}
  \caption{\label{fig:selection}Charge--spin selection rules and harmonic separation in odd-parity magnets. Contributions of response order $N=1,2,3,4,\dots$ to output harmonics $n\Omega$ under monochromatic driving $E(t)=E_0\cos\Omega t$. Cyan (salmon) cells denote symmetry-allowed charge (spin) contributions. Crossed and gray cells denote zero contributions: crosses indicate opposite parities of $N$ and $n$, while gray cells indicate $n>N$. For a monochromatic drive, an $N$th-order response can generate harmonics only up to $N\Omega$. Multiple response orders can contribute to the same output harmonic, while charge and spin occupy odd and even harmonics, respectively, including dc rectified spin current.}
\end{figure}

In this Letter, we establish temporal and spectral charge--spin separation as the defining transport signature of OPMs of all spin-splitting classes. This is the dynamical counterpart of spatial spin separation in even-parity altermagnets. For collinear and noncollinear OPMs protected by $\Uop=[C_2^\perp\Vert P]$, we prove that charge and $z$-polarized spin currents are odd and even under electric-field reversal, respectively. The resulting mechanism-independent selection rules allow only odd-order charge responses and even-order spin responses, without assuming spin conservation. Under a half-period antisymmetric ac drive, the charge current reverses sign after $T/2$ while the spin current repeats, yielding odd charge and even spin harmonics. These complementary harmonics allow dc spin rectification without a dc charge counterpart, including in longitudinal geometry. The harmonic separation persists when multiple response orders mix in strong fields. In addition, we verify the selection rules for all linear and second-order transport mechanisms and show that antiunitary $\mathcal T\tau$ also enforces odd-parity spin splitting but reverses the charge--spin channel selection for Berry-curvature and Berry-curvature-dipole responses. We demonstrate the charge--spin selection rules and temporal--spectral separation across a minimal odd-parity model and two representative material platforms: Floquet-engineered collinear MnPS$_3$ and the equilibrium noncollinear VCl$_2$ family. Figure~\ref{fig:selection} illustrates the resulting all-order selection rules and harmonic separation.

\textit{All-order charge--spin selection rules---}\phantomsection\label{sec:all-order}We consider collinear and noncollinear OPMs whose Hamiltonians are invariant under the unitary spin-flip inversion symmetry $\Uop=[C_2^\perp\Vert P]$. Here $C_2^\perp$ is a $\pi$ spin rotation about an axis perpendicular to the spin-polarization axis $z$ and $P$ is spatial inversion. This spin--group symmetry $\Uop$ is a key unitary mechanism for protecting odd-parity spin splitting across collinear and noncollinear magnetic orders~\cite{zengOddParitySpinGroup2026,luoSpinGroupCriteria2026,huangLightInducedOddParity2026,zhuFloquetOddParity2026}. It relates equal-energy states at opposite momenta with $S_{z,\nu}(-\bk)=-S_{z,\nu'}(\bk)$, where $\nu'$ labels the symmetry-related partner band and $S_{z,\nu}=\langle u_{\nu\bk}|\hat s_z|u_{\nu\bk}\rangle$.
In a nonzero driving electric field $\bE$, this zero-field symmetry maps the driven Hamiltonian between opposite field configurations:
\begin{equation}
  {\Uop} {\hat H}[\bE]{\Uop}^{-1}={\hat H}[-\bE].
  \label{eq:H-covariance}
\end{equation}
In generic dissipative transport with $\Uop$-preserving scattering and boundary conditions, as naturally satisfied by symmetric scattering and the standard relaxation-time approximation (RTA)~\cite{peresOpticalBistability2014}, relaxation in a driven electric field $\bE$ selects a unique steady state $\hat\rho_{\bE}$ inheriting the covariance of Eq.~\eqref{eq:H-covariance}, yielding ${\Uop}{\hat\rho}_{\bE}{\Uop}^{-1}={\hat\rho}_{-\bE}$. A general Liouvillian proof is provided in the Supplemental Material (SM)~\cite{supplemental}.

The charge-current and conventional $z$-polarized spin-current operators are
\begin{equation}
  \hat J_{c,i}=-e\hat v_i,\qquad  \hat J_{s_z,i}=-\frac{e}{\hbar} \{\hat v_i,\hat s_z\},
  \label{eq:two-current}
\end{equation}
with both currents expressed in charge-equivalent units~\cite{ShiProperSpinCurrent2006}. Here $\hat v_i$ and $\hat s_z$ are the velocity and spin operators. Since $\Uop$ reverses both $\hat v_i$ and $\hat s_z$, the charge-current operator is odd under $\Uop$, whereas the spin-current operator is even. Writing the field-induced currents as $\bm J_{\alpha}(\bE)=\operatorname{Tr}({\hat \rho}_{\bE}\hat{\bm J}_{\alpha})$, with $\alpha\in\{c,s_z\}$ denoting the charge and spin channels, and using ${\Uop}{\hat \rho}_{\bE}{\Uop}^{-1}={\hat \rho}_{-\bE}$, we obtain the two central relations
\begin{equation}
  \bm J_c(\bE)=-\bm J_c(-\bE),\qquad
  \bm J_{s_z}(\bE)=+\bm J_{s_z}(-\bE).
  \label{eq:master-relations}
\end{equation}
Expanding in powers of the electric field,
\begin{equation}
  J_{\alpha,i}(\bE)=\sum_{N\geq1}
  \chi_{ia_1\cdots a_N}^{\alpha,(N)}E_{a_1}\cdots E_{a_N},
  \label{eq:response-expansion}
\end{equation}
and comparing both sides of Eq.~\eqref{eq:master-relations} order by order yields
\begin{equation}
  \chi_c^{(2m)}=0,\qquad
  \chi_{s_z}^{(2m+1)}=0,
  \label{eq:all-order-selection}
\end{equation}
with $m\geq1$ ($m\geq0$) for the charge (spin) channel. Equation~\eqref{eq:master-relations} depends only on how the driven state and the current operators transform under $\Uop$. The resulting all-order selection rules therefore apply to collinear and noncollinear OPMs of every odd-parity spin-splitting class.

\textit{Harmonic separation theorem---}\phantomsection\label{sec:dynamical}We now extend the all-order selection rules to nonperturbative periodic transport using dynamical symmetry. For time-dependent electric fields, $\Uop$ promotes to a dynamical symmetry that directly constrains  the nonperturbative harmonic spectrum~\cite{WernerAltermagnetHHG2024,neufeldFloquetGroupTheory2019,EngelhardtDynamical2021,mizuno_highharmonic_2026a,IkedaSpinHHG2019,ZhangMagneticHHG2018,TakayoshiSpinHHG2019}. For any $T$-periodic drive satisfying $\bE(t+T/2)=-\bE(t)$, Eq.~\eqref{eq:H-covariance} gives the dynamical symmetry $\Dop=(\Uop,T/2)$. Monochromatic fields with linear, circular, or elliptical polarization satisfy this condition, as do multitone drives containing only odd harmonics of the fundamental. The periodic steady state therefore satisfies ${\Uop}{\hat \rho}(t){\Uop}^{-1}={\hat \rho}(t+T/2)$.
The opposite parities of the two current operators under $\Uop$ immediately imply
\begin{equation}
  J_c(t+T/2)=-J_c(t),\qquad
  J_{s_z}(t+T/2)=+J_{s_z}(t).
  \label{eq:half-period-currents}
\end{equation}
In the time domain, Eq.~\eqref{eq:half-period-currents} dictates that $J_c(t)$ reverses sign every half-cycle while $J_{s_z}(t)$ repeats with period $T/2$. In the frequency domain, Fourier transforming Eq.~\eqref{eq:half-period-currents} yields
\begin{equation}
  J_c^{2m\Omega}=0,\qquad
  J_{s_z}^{(2m+1)\Omega}=0.
  \label{eq:harmonic-selection}
\end{equation}
Here $m\geq0$ in both channels, and $J_\alpha^{n\Omega}\equiv T^{-1}\int_0^T J_\alpha(t)e^{in\Omega t}\,\mathrm dt$ is the $n$th Fourier coefficient, where $\Omega=2\pi/T$ is the frequency of the drive. Thus charge currents contain only odd harmonics, whereas spin currents contain only even harmonics, including a pure dc spin current free of any dc charge counterpart.
 
The harmonic index $n$ (output frequency $n\Omega$) is distinct from the perturbative response order $N$ (power of the field amplitude $E_0$). Under a monochromatic drive, each allowed harmonic resums all response orders of the same parity, $N=n, n+2, \dots$:
\begin{equation}
  J_{\alpha}^{n\Omega}=\sum_{l=0}^{\infty}C_{\alpha,n+2l}^{n\Omega}E_0^{n+2l}.
  \label{eq:harmonic-expand}
\end{equation}
For $n = 0$, the field-induced dc spin current starts at second order. Although higher-order terms ($l\ge 1$) modify the weak-field $E_0^n$ scaling in the strong-field regime, the dynamical symmetry in Eq.~\eqref{eq:half-period-currents} guarantees that the separation into odd charge and even spin harmonics remains exact beyond the weak-field regime.

\textit{Symmetry constraints on specific microscopic transport mechanisms---}\phantomsection\label{sec:mechanisms}To verify the mechanism independence of the selection rules, we evaluate the complete first- and second-order charge and spin responses without assuming spin conservation. The charge and spin current vertices are ${\hat w}_i^c={\hat v}_i$ and ${\hat w}_i^{s_z}=\{{\hat v}_i,{\hat s_z}\}/\hbar$, with $\hat J_{\alpha,i}=-e{\hat w}_i^\alpha$. We expand the density operator, Hamiltonian, and current operator consistently in the driving field, retaining all contact terms. The complete finite-frequency expressions are given in Eqs.~\eqref{eq:em-linear-response} and~\eqref{eq:em-second-response} of the End Matter (see SM~\cite{supplemental} for detailed derivations).

Using the spectral decomposition of the Green's functions, we express each response as a sum of terms $\mathcal W X$. Here $X$ is a trace of products of vertices and band projectors, while the scalar coefficient $\mathcal{W}$ depends on the energies, occupations, frequencies, and broadening. Under $\Uop$, every structure at order $N=1,2$, including all contact terms, obeys
\begin{equation}
  X_{\Uop\mathcal N}^{\alpha}(-\bk)
    =q_\alpha^{(N)}X_{\mathcal N}^{\alpha}(\bk),\qquad
  q_\alpha^{(N)}=\eta_\alpha(-1)^N,
  \label{eq:micro-common-sign}
\end{equation}
where $\mathcal N$ collects band indices, $\eta_c=-1$, and $\eta_{s_z}=+1$. Symmetry partners have identical ${\mathcal W}$, so their combined contribution is $(1+q_\alpha^{(N)}){\mathcal W}X$. Each term in the linear spin response and the second-order charge response therefore vanishes after summing over symmetry-related partners. This cancellation holds for the full complex response before taking a dc or clean limit, irrespective of spin mixing or lifetime scaling. It explicitly verifies the mechanism independence of the all-order selection rules above.

The antiunitary symmetry $\mathcal A=\mathcal T\tau$, with magnetic fractional translation $\tau$, produces the odd-parity band relation and can lead to charge--spin separation for specific transport mechanisms. However, its antiunitarity maps retarded Green's functions to advanced ones, $\mathcal A\hat G^R(\epsilon,\bk)\mathcal A^{-1}=\hat G^A(\epsilon,-\bk)$, so the unitary steady-state covariance argument cannot be applied directly. To compare symmetry-related contributions to the same causal response, we keep the retarded prescription fixed. The weights $\mathcal{W}$ are then identical, whereas antiunitarity complex conjugates the matrix products: $X_{\mathcal A\mathcal N}^{\alpha}(-\bk)=q_\alpha^{(N)}[X_{\mathcal N}^{\alpha}(\bk)]^*$. Consequently, in $\mathcal A$-protected systems, contributions governed by $\operatorname{Re}X$ obey the same selection rules as under $\Uop$, whereas those governed by $\operatorname{Im}X$ follow the opposite rules (Table~\ref{tab:full-response-selection}). Because both $\operatorname{Re}X$ and $\operatorname{Im}X$ carry the full complex weight $\mathcal W$, $\mathcal A$ alone cannot enforce a universal charge--spin separation across all mechanisms.

In the spin-conserving, clean dc limit, the response reduces to the six well-established mechanisms: linear Drude and Berry curvature (BC), and second-order nonlinear Drude, Berry curvature dipole (BCD), and interband and intraband quantum metric dipoles (QMD)~\cite{sundaramWavePacket1999,sodemannBerryCurvatureDipole2015,gaoFieldInducedPositionalShift2014,ulrichQuantumGeometric2026,shibataSecondOrderKeldysh2026,KaplanQuantumGeometry2024,WangIntrinsicHall2021,LiuIntrinsicHall2021,DuQuantumNonlinearHall2021}. Under $\Uop$, both linear mechanisms forbid spin current and all four second-order mechanisms forbid charge current. Under $\mathcal A$, Drude, nonlinear Drude, and both QMD terms follow the real-structure rule, while BC and BCD follow the imaginary-structure rule, allowing linear spin BC and second-order charge BCD responses.

Table~\ref{tab:mechanisms-comparison} compares these results with the intrinsic symmetry $\Theta=[C_2^\perp\mathcal T\Vert E]$ of altermagnets~\cite{smejkalBeyondConventional2022}, where $E$ denotes the spatial identity. Unlike $\Uop$ and $\mathcal A$, $\Theta$ preserves each spin block: it forbids BC, nonlinear Drude, and both QMD terms in both channels, while allowing Drude and BCD subject to the remaining magnetic-group constraints. Considered individually (Table~\ref{tab:mechanisms-comparison}), $\Uop$ uniquely enforces complete channel separation across all mechanisms, whereas $\mathcal A$ and $\Theta$ leave distinct fingerprints in the BC and BCD channels, offering a direct transport diagnostic to distinguish unitary OPMs, antiunitary OPMs, and altermagnets.

\textit{Nonperturbative periodic transport method---}\phantomsection\label{sec:nonperturbative}At first and second order, symmetry-related contributions cancel in each forbidden channel for both intraband and interband processes (Tables~\ref{tab:full-response-selection} and~\ref{tab:mechanisms-comparison}). Under a strong monochromatic drive, several orders $N=n,n+2,\ldots$ can contribute appreciably to a single harmonic $n\Omega$. To capture this mixing, we use Chambers' method to solve the time-dependent intraband Boltzmann equation within the RTA, retaining the full field dependence~\cite{chambersKineticFormulation1952,buddChambersSolution1962,maccallumKineticIntegral1963}.

For $E(t)=E_0\cos\Omega t$ along $x$, the occupation of spinor band $\nu$ obeys
\begin{equation}
  \partial_t f_\nu
  -\frac{eE(t)}{\hbar}\partial_{k_x}f_\nu
  =-\frac{f_\nu-f^{(0)}_\nu}{\tau_r}.
  \label{eq:rta}
\end{equation}
Here $f_\nu^{(0)}$ is the reference Fermi--Dirac distribution and $\tau_r$ is the relaxation time. The resulting field-induced current is
\begin{equation}
  \begin{aligned}
    J_{\alpha,i}(t)={}&-\frac{e}{\tau_r}\int_0^\infty\!\mathrm du\,e^{-u/\tau_r}
    \sum_{\nu,\bk}\zeta_{i\nu}^{\alpha}(\bk)\\
    &\times \{f_\nu^{(0)}[\bk+q(t,u)\hat{\bm x}]-f_\nu^{(0)}(\bk)\},
  \end{aligned}
  \label{eq:periodic-current}
\end{equation}
where $\alpha\in\{c,s_z\}$, $u\geq0$ is the elapsed memory time, and $q(t,u)=(e/\hbar)\int_{t-u}^{t}E(l)\,\mathrm dl$. For the monochromatic drive, $q(t,u)=K_0[\sin\Omega t-\sin\Omega(t-u)]$, with $K_0=eE_0/(\hbar\Omega)$. With the equilibrium background subtracted, Eq.~\eqref{eq:periodic-current} retains all field-amplitude orders and finite-frequency phase lags within the intraband RTA (see SM Secs. S3 and S4~\cite{supplemental}). The charge and spin velocity kernels are $\zeta_{i\nu}^{c}=\hbar^{-1}\partial_{k_i}\varepsilon_\nu$ and $\zeta_{i\nu}^{s_z}=\hbar^{-1}\langle u_{\nu\bk}\vert{}\{\hat v_i,\hat s_z\}\vert{}u_{\nu\bk}\rangle$. For conserved $\hat s_z$, the spin kernel reduces to $\pm v_{i\nu}$. In noncollinear OPMs the spin kernel retains its full anticommutator form and remains even under $\Uop$. This even parity, together with the odd charge kernel and the half-period reversal of the drive, yields Eq.~\eqref{eq:harmonic-selection}. We use Eq.~\eqref{eq:periodic-current} to compute the current waveforms for the model and materials below and obtain their harmonics by Fourier transformation.

\begin{figure}[t]
  \centering
  \includegraphics[width=\columnwidth]{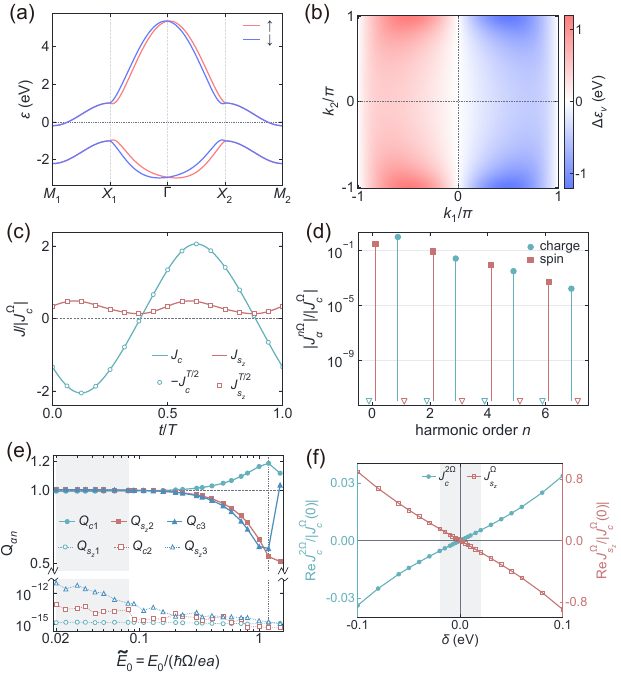}
  \caption{\label{fig:model}
  Nonperturbative harmonic separation in the $p$-wave model. (a) Spin-resolved bands of the two-sublattice, four-band model with conserved ${\hat s_z}$. (b) Odd-parity spin splitting $\Delta\varepsilon_v(\bk)$ of the valence band. (c) Time-domain waveforms of the charge current $J_c$ and spin current $J_{s_z}=J_\uparrow-J_\downarrow$, together with the same waveforms shifted by half a period ($J_\alpha^{T/2}\equiv J_\alpha(t+T/2)$), normalized by $|J_c^\Omega|$. (d) Fourier amplitudes $|J_\alpha^{n\Omega}|$ for $n=0,\ldots,7$ at $\widetilde E_0=1.2$; open triangles mark forbidden components at the numerical floor. (e) Field-amplitude scan: allowed harmonics depart from their weak-field scaling $|J_\alpha^{n\Omega}|\propto\widetilde E_0^n$ in strong fields, while forbidden components remain numerically zero. At each harmonic order, the amplitudes in both channels are normalized by the same weak-field reference for the allowed channel. The vertical dotted line marks $\widetilde E_0=1.2$, used in (d). (f) $\operatorname{Re}J_c^{2\Omega}$ and $\operatorname{Re}J_{s_z}^{\Omega}$ induced by the $\Uop$-breaking term $\delta\tau_z$, normalized by $|J_c^\Omega(\delta=0)|$. Parameters are $t=\Delta_z=1.0~\mathrm{eV}$, $t_0=0.3~\mathrm{eV}$, and $\lambda_p=0.3~\mathrm{eV}$, with $\mu=-3.0~\mathrm{eV}$, $T_{e}=300~\mathrm{K}$ and $\Omega\tau_r=1$. We set $\delta=0$ except in (f).}
\end{figure}

\textit{Minimal odd-parity model---}\phantomsection\label{sec:model}The selection rules and harmonic separation apply to every odd-parity spin-splitting class. A minimal example is a collinear $p$-wave lattice model with two sublattices and two conserved spin blocks. Its four-band Hamiltonian is
\begin{equation}
  \begin{aligned}
    H_\delta(\bk)={}&d_0(\bk)\tau_0+d_x(\bk)\tau_x\\
    &+[d_p(\bk)+\delta]\tau_z+\Delta_z\tau_z\sigma_z,
  \end{aligned}
  \label{eq:model-H}
\end{equation}
where $d_0=2t_0(\cos k_1+\cos k_2)$, $d_x=4t\cos\frac{k_1}{2}\cos\frac{k_2}{2}$, and $d_p=2\lambda_p\sin k_1$, with Pauli matrices $\tau_i$ and $\sigma_i$ acting on the sublattice and spin spaces. At $\delta=0$, spin-flip inversion $U_{\Uop}=\tau_x\sigma_x$ protects an exact odd-parity spin splitting $\Delta\varepsilon_\nu(-\bk)=-\Delta\varepsilon_\nu(\bk)$ [$\Delta\varepsilon_\nu(\bk)=\varepsilon_{\nu,\uparrow}(\bk)-\varepsilon_{\nu,\downarrow}(\bk)$ for band $\nu=\pm$], as shown in Figs.~\ref{fig:model}(a) and \ref{fig:model}(b).

\begin{figure*}[!t]
  \centering
  \includegraphics[width=\textwidth]{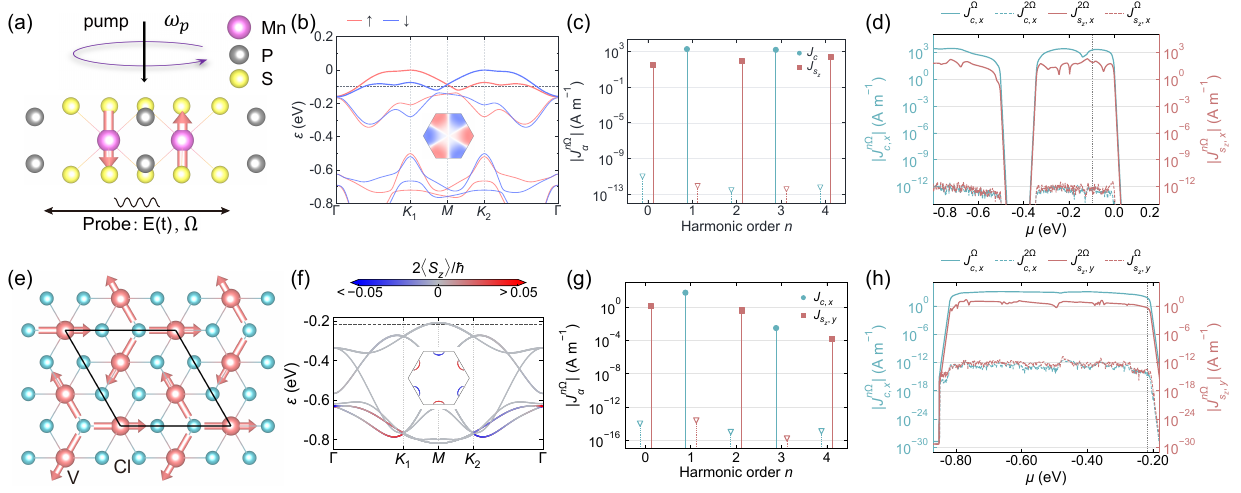}
  \caption{\label{fig:material}
  Charge--spin harmonic separation in collinear and noncollinear odd-parity magnet materials. (a)--(d) Floquet monolayer MnPS$_3$. (a) Structure and pump--probe scheme: purple, gray, and yellow spheres denote Mn, P, and S, respectively, and arrows show the collinear N\'eel order. (b) Floquet bands with spin up (red) and down (blue); inset: pump-induced spin splitting of the highest valence band. (c) Harmonics $n=0,\ldots,4$ of the longitudinal charge current $J_{c,x}$ and spin current $J_{s_z,x}$. (d) Fundamental and second harmonics versus chemical potential. Energies are measured relative to the Floquet valence-band maximum; the horizontal dashed line in (b) and vertical dotted line in (d) mark the filling used in (c). (e)--(h) Noncollinear monolayer VCl$_2$. (e) Coplanar $120^\circ$ antiferromagnetic order. (f) Valence bands colored by $2\langle {\hat s_z}\rangle/\hbar$; inset: spin-polarized Fermi contours. (g) Harmonic amplitudes of the longitudinal charge current $J_{c,x}$ and transverse spin current $J_{s_z,y}$. (h) Fundamental and second harmonics versus chemical potential; the vertical dotted line marks the filling in (g). For VCl$_2$, energies are measured relative to $E_F$, and the horizontal dashed line in (f) marks the filling used in (g). In (c),(g), filled circles (squares) denote allowed charge (spin) components, and open inverted triangles denote forbidden residuals. In (d),(h), solid (dashed) curves denote allowed (forbidden) components, with charge and spin amplitudes on the left and right axes, respectively. Spin currents use charge-equivalent units.}
\end{figure*}

Figures~\ref{fig:model}(c) and ~\ref{fig:model}(d) show nonperturbative harmonic separation in both the time and frequency domains. In the time domain, the waveforms exhibit distinct temporal symmetries, $J_c(t+T/2)=-J_c(t)$ versus $J_{s_z}(t+T/2)=+J_{s_z}(t)$ [Fig.~\ref{fig:model}(c)], visually demonstrating temporal separation between a period-$T$ charge current and a period-$T/2$ spin current. We define the dimensionless field amplitude $\widetilde E_0=aeE_0/(\hbar\Omega)$, where $a$ is the lattice constant. In the Fourier spectrum [Fig.~\ref{fig:model}(d)], charge currents appear exclusively at odd harmonics ($n=1,3,5,7$) and spin currents at even harmonics ($n=0,2,4,6$), with $\vert{}J_{s_z}^{0\Omega}\vert{}$ comparable in magnitude to $\vert{}J_c^\Omega\vert{}$ and all forbidden harmonics suppressed to numerical precision.

Figure~\ref{fig:model}(e) tests the breakdown of weak-field scaling using $Q_{\alpha n}=|J_\alpha^{n\Omega}|/
(C_n^{\mathrm{ref}}\widetilde E_0^n)$, where $C_n^{\mathrm{ref}}$ is fitted to the allowed harmonic at weak fields and shared by both channels. As $\widetilde E_0$ increases beyond the perturbative regime, higher-order mixing drives the allowed harmonics away from $Q_{\alpha n}=1$, whereas all forbidden components remain at the numerical floor. Conversely, introducing a $\Uop$-breaking staggered potential $\delta\tau_z$ activates $J_{s_z}^{\Omega}$ and $J_c^{2\Omega}$ [Fig. ~\ref{fig:model}(f)]. Since $U_{\Uop} H_\delta(\bk)U_{\Uop}^{-1}=H_{-\delta}(-\bk)$, both induced harmonics are odd in $\delta$, reversing sign across $\delta=0$ with a linear onset that directly probes $\Uop$ breaking~\cite{EvenTzurBrokenSelection2022}. The same criterion applies to SOC: terms preserving $\Uop$ retain exact harmonic separation despite spin mixing, whereas terms breaking $\Uop$ activate forbidden harmonics (see SM~\cite{supplemental}).

\textit{Material realization---}\phantomsection\label{sec:material}We next demonstrate harmonic separation in two realistic materials: Floquet-engineered collinear monolayer MnPS$_3$ and equilibrium noncollinear monolayer VCl$_2$. We apply a strong probe field to Floquet MnPS$_3$ and a weak ac field to equilibrium VCl$_2$. Together, they test the $\Uop$-enforced selection rules across complementary regimes: a strong-field, spin-conserving $f$-wave Floquet state with longitudinal spin current in MnPS$_3$, and a weak-field, spin-nonconserving equilibrium state with transverse spin current in VCl$_2$.

Motivated by established antiferromagnetism in bulk and few-layer MnPS$_3$~\cite{sunProbingMnPS32019,niAntiferromagneticDomainsMnPS32021} and the experimental isolation of freestanding monolayers~\cite{kinyanjuiFreestandingMnPS32018}, we study monolayer MnPS$_3$ with collinear N\'eel order on the Mn honeycomb lattice. A high-frequency circularly polarized pump breaks $P\mathcal T$ and induces $f$-wave odd-parity spin splitting in the effective Floquet bands while preserving $\mathcal U$ and the conservation of $\hat s_z$ [Fig.~\ref{fig:material}(a,b)]~\cite{huangLightInducedOddParity2026,zhuFloquetOddParity2026,liuLightInducedDimerized2026,zhangTriangulene2026,li_floquet_2026,liu_complete_2026}. Driving the dressed bands with a lower-frequency ac field $\bE_{\rm pr}(t)=E_0\cos\Omega t\,\hat{\bm x}$ ($\Omega\ll\omega_p$, with $\omega_p$ the Floquet frequency), we evaluate the nonperturbative intraband response within the RTA using Eq.~\eqref{eq:periodic-current}. Figures~\ref{fig:material}(c,d) confirm odd charge and even spin harmonics in this multiband collinear realization: filling tunes the allowed amplitudes, while forbidden components remain suppressed throughout the scanned chemical-potential range.

Coplanar $120^\circ$ antiferromagnetic order has been predicted for centrosymmetric $1T$ transition-metal dihalide monolayers $MX_2$ ($M=\mathrm{V,Mn}$; $X=\mathrm{Cl,Br}$)~\cite{nguyenExchangeRules2024,sodequistSpinSpiralHalides2023}; for VCl$_2$, this order has been established experimentally in bulk, and two-dimensional flakes have been synthesized~\cite{kadowakiVCl2PhaseTransition1987,jiangDihalideSynthesis2023}. We identify these monolayer magnetic phases as a family of noncollinear OPMs protected by $\Uop$ in the nonrelativistic limit. Taking VCl$_2$ as a representative member [Fig.~\ref{fig:material}(e)], its magnetic structure preserves $\Uop=[C_{2x}\Vert P]$ and $[E\Vert C_{2y}]$. For the $x$-directed drive, $[E\Vert C_{2y}]$ and $\Uop$ enforce $J_{c,y}=J_{s_z,x}=0$, leaving longitudinal charge ($J_{c,x}$) and transverse $z$-polarized spin ($J_{s_z,y}$) currents symmetry-allowed. Although the bands exhibit weak out-of-plane spin polarization [Fig.~\ref{fig:material}(f)] and nonconserved $\hat s_z$, the full spin-current operator remains constrained by $\Uop$. The resulting currents from Eq.~\eqref{eq:periodic-current} display complementary harmonic spectra, featuring a pure transverse dc spin current with no dc charge counterpart [Fig.~\ref{fig:material}(g,h)]. As mobile holes are depleted, the allowed signals decrease while forbidden components remain at numerical residual levels. 

While Fig.~\ref{fig:material} shows the nonrelativistic results, incorporating first-principles SOC preserves the dominant odd-charge and even-spin harmonic hierarchy in both materials: forbidden harmonics remain suppressed to the percent level in MnPS$_3$ and by more than four orders of magnitude in VCl$_2$ (see SM~\cite{supplemental}).


\textit{Discussion---}In collinear and noncollinear OPMs of any odd-parity wave, unitary spin-flip inversion $\Uop=[C_2^\perp\Vert P]$ restricts charge responses to odd orders and $z$-polarized spin responses to even orders, irrespective of the microscopic mechanism.

Like altermagnetism, odd-parity magnetism is defined within the spin-group framework in the nonrelativistic limit~\cite{smejkalBeyondConventional2022,hellenesPWaveMagnets2023,zengOddParitySpinGroup2026,luoSpinGroupCriteria2026}. As in altermagnets, the persistence of spin-group-protected transport beyond this limit hinges on how SOC acts on the protecting symmetry; our model analysis and first-principles calculations show that the selection rules remain exact for $\Uop$-preserving SOC despite strong spin mixing, and that the complementary charge--spin harmonic pattern remains well resolved even under weak SOC-induced symmetry breaking (see details in SM~\cite{supplemental}). In contrast to the crystal spin Hall effect, which is restricted to $d$-wave ($l=2$) altermagnets~\cite{gonzalezHernandezElectricalSpinSplitter2021,douAnisotropicSpinPolarized2025}, the charge--spin selection rules and temporal--spectral separation established here are governed by unitary spin-flip inversion rather than a particular angular form of spin splitting, and therefore extends across all $\Uop$-protected odd-parity classes ($l=1,3,\cdots,9$).

Experimentally, a single monochromatic drive already satisfies the half-period antisymmetry condition. In collinear platforms, a circularly polarized pump paired with a lower-frequency probe combines optical control of odd-parity bands with harmonic-resolved electrical and spin readout; in equilibrium noncollinear platforms, an in-plane ac drive enables direct lock-in detection of odd charge harmonics alongside dc and even-harmonic spin signals via inverse spin Hall or magneto-optical probes.

The charge--spin selection rules and harmonic separation theorem proposed here have three implications. (i) The complementary spectra enable ac-driven spin rectification and spin-harmonic generation, with dc and even-harmonic spin currents having no charge counterparts at the same frequencies, even in longitudinal geometries. (ii) While carrier filling and symmetry-preserving scattering can tune the allowed responses, the emergence of forbidden harmonics provides a sensitive probe of symmetry breaking, including that induced by SOC. (iii) The characteristic charge--spin selection rules and harmonic separation provide a transport fingerprint of OPMs, complementary to spin- and angle-resolved photoemission spectroscopy. In particular, although both $\Uop$ and $\mathcal T\tau$ can protect odd-parity spin splitting, they impose opposite charge--spin selection rules for Berry-curvature and Berry-curvature-dipole responses. Linear and nonlinear Hall measurements can thus distinguish the two protecting symmetries. Our results therefore establish temporal and spectral charge--spin separation as a distinct transport paradigm complementary to the spatial separation characteristic of even-parity altermagnets.

\begin{acknowledgments}
\textit{Acknowledgments---}The work is supported by Science Fund for Creative Research Groups of NSFC (Grant No. 12321004), the NSF of China (Grant No. 12374055), and the Fundamental Research Funds for the Central Universities (Grant No. 2026CX01001).
\end{acknowledgments}

\nocite{bucaSymmetryLindblad2012,marcianiBoltzmannTransport2021,spohnAlgebraicCondition1977,sundaramWavePacket1999,sodemannBerryCurvatureDipole2015,gaoFieldInducedPositionalShift2014,ulrichQuantumGeometric2026,shibataSecondOrderKeldysh2026,GuoDissipationGeometry2026,ParkerDiagrammatic2019,jiaNonlinearMagnetoelectric2026,zhangNonlinearChargeSpin2021,zhu_magnetic_2025,venturaGaugeCovariances2017,langrethResponseTheory1976,haugQuantumKinetics2008,maccallumKineticIntegral1963,peresOpticalBistability2014,kresseEfficientIterative1996,blochlProjectorAugmented1994,kresseUltrasoftPAW1999,perdewGeneralizedGradient1996,dudarevLSDAU1998,liechtensteinDensityFunctional1995,pizziWannier902020,huangLightInducedOddParity2026,Kosugi2014}
\bibliography{references}

\clearpage
\appendix
\onecolumngrid
{\section*{End matter}}
\phantomsection\label{sec:end-matter}

\subsection*{Complete charge and spin responses up to the second order at finite frequency}
For uniform fields and bulk currents, we use independent quasiparticles with Peierls coupling $\hat H_{\bm A}(\bk)=\hat H_0(\bk+e\bm A/\hbar)$ and field-independent scalar broadening $\gamma=\hbar/(2\tau_r)>0$, common to symmetry partners. The charge-current construction~\cite{shibataSecondOrderKeldysh2026} extends to nonconserved spin by retaining the full measurement and contact vertices,
\begin{equation}
  \begin{gathered}
    \hat w_i^c=\hat v_i,\qquad \hat w_i^{s_z}=\frac{1}{\hbar}\{\hat v_i,\hat s_z\},\qquad
    \hat v_j=\frac{1}{\hbar}D_j\hat H_0,\\
    \hat V_{jk}=\frac{1}{\hbar}D_j\hat v_k,\qquad
    \hat w_{i;j}^\alpha=\frac{1}{\hbar}D_j\hat w_i^\alpha,\qquad
    \hat w_{i;jk}^\alpha=\frac{1}{\hbar^2}D_jD_k\hat w_i^\alpha.
  \end{gathered}
  \label{eq:em-vertices}
\end{equation}
Here $D_j$ is the physical momentum derivative, covariant under band-basis changes, with $D_j\hat s_z=0$; its definition and the full derivation are given in SM Sec.~S2.1~\cite{supplemental}. Expanding both the driven state and measured current retains all driving and measurement contact terms. With $E_j(\omega)=i\omega A_j(\omega)$, $\hat G_a=\hat G(\epsilon+\hbar\omega_a)$, $\hat G_0=\hat G(\epsilon)$, and $\omega_{12}=\omega_1+\omega_2$, take $\hat G^{R/A}(\epsilon)=[\epsilon-\hat H_0\pm i\gamma]^{-1}$ and $\hat G^<(\epsilon)=f_{\rm FD}(\epsilon)[\hat G^A(\epsilon)-\hat G^R(\epsilon)]$. The complete first- and second-order responses are
\begin{equation}
  \chi_{ij}^{\alpha,(1)}(\omega)
    =\frac{e^2}{\omega}\int_{\bk}\int\frac{d\epsilon}{2\pi}
      \left\{\operatorname{Tr}[\hat w_i^\alpha \hat G_\omega \hat v_j\hat G_0]^<
          +\operatorname{Tr}[\hat w_{i;j}^\alpha \hat G_0^<]\right\},
  \label{eq:em-linear-response}
\end{equation}
and
\begin{equation}
  \begin{aligned}
    \chi_{ijk}^{\alpha,(2)}(\omega_1,\omega_2)
      ={}&-\frac{ie^3}{2\omega_1\omega_2}
        \int_{\bk}\int\frac{d\epsilon}{2\pi}
        \Big\{\operatorname{Tr}[\hat w_i^\alpha \hat G_{12}\hat v_j\hat G_2\hat v_k\hat G_0]^<
             +\operatorname{Tr}[\hat w_i^\alpha \hat G_{12}\hat v_k\hat G_1\hat v_j\hat G_0]^<\\
      &\quad+\operatorname{Tr}[\hat w_i^\alpha \hat G_{12}\hat V_{jk}\hat G_0]^<
             +\operatorname{Tr}[\hat w_{i;j}^\alpha \hat G_2\hat v_k\hat G_0]^<
             +\operatorname{Tr}[\hat w_{i;k}^\alpha \hat G_1\hat v_j\hat G_0]^<
             +\operatorname{Tr}[\hat w_{i;jk}^\alpha \hat G_0^<]\Big\}.
  \end{aligned}
  \label{eq:em-second-response}
\end{equation}
Here $\int_{\bk}=\int_{\rm BZ}d^dk/(2\pi)^d$ and $<$ denotes the lesser projection of the entire ordered product. The second-order {response} tensor is symmetrized under $(j,\omega_1)\leftrightarrow(k,\omega_2)$. Currents exclude their equilibrium background. Dyson and lesser-component expansions, diagrams, and frequency convolutions are provided in SM Sec.~S2.1~\cite{supplemental}.

\subsection*{Symmetry constraints on the response terms}
Resolving the Green functions into projectors gives $\chi_\alpha^{(N)}=\sum_{a,\mathcal N}\int_{\bk}\mathcal W_{\mathcal N}^{a,(N)}X_{\mathcal N}^{a,\alpha}$ for $N=1,2$. Here $X$ is a closed vertex product, such as $\operatorname{Tr}(\hat P_n\hat w_i^\alpha \hat P_m\hat v_j\hat P_l\hat v_k)$, and $\mathcal W$ contains its scalar propagation weights and energy integral; $a$ labels structures and input orderings, and $\mathcal N$ collects band indices. The complete list of structures and weights is given in SM Sec.~S2.1~\cite{supplemental}.

For $\Uop$ and $\mathcal A=\mathcal T\tau$, each driving velocity or physical momentum derivative contributes a minus sign. Together with $\eta_c=-1$ and $\eta_{s_z}=+1$, this gives $q=\eta_\alpha(-1)^N$ for every structure, including contact terms. The partner product is $qX$ under $\Uop$ and $qX^*$ under $\mathcal A$; translation phases cancel. Because we compare symmetry-partner contributions within the same causal (retarded) response function rather than conjugating $\hat G^R$ into $\hat G^A$, equal partner energies, occupations, and broadenings guarantee identical weights $\mathcal W_{\mathcal N}(-\bk)=\mathcal W_{\mathcal N}(\bk)$. Thus
\begin{equation}
  \begin{aligned}
    \Uop:\quad \mathcal WX+\mathcal W_{\Uop}X_{\Uop}&=(1+q)\mathcal WX,\\
    \mathcal A:\quad \mathcal WX+\mathcal W_{\mathcal A}X_{\mathcal A}
      &=\mathcal W\big[(1+q)\operatorname{Re}X+i(1-q)\operatorname{Im}X\big].
  \end{aligned}
  \label{eq:em-partner-sums}
\end{equation}
Under $\Uop$, $q=-1$ eliminates the entire complex contribution after partner summation. Under $\mathcal A$, real and imaginary matrix structures select opposite channels (Table~\ref{tab:full-response-selection}); these are not the real and imaginary parts of $\chi$, since $\mathcal W$ is complex. The more general all-order selection rules and harmonic separation theorem follow from dynamical covariance under the conditions stated in the main text; see SM Sec.~S1~\cite{supplemental}.

\newpage
\begin{table}[!htbp]
  \caption{\label{tab:full-response-selection}
  Channel constraints on the complete first- and second-order matrix structures, without assuming conserved spin. Each part of $X$ retains its full complex propagation weight. Parities $(\pi_c,\pi_{s_z})$ refer to the indicated matrix part. The symmetries are imposed separately; $\checkmark$ ($\times$) denotes a symmetry-allowed (forbidden) contribution under the specified operation alone.}
  \centering\small
  \renewcommand{\arraystretch}{1.22}
  \setlength{\tabcolsep}{8pt}
  \begin{tabular}{@{}cc*{2}{ccc}@{}}
    \toprule
    & & \multicolumn{3}{c}{$\Uop$} & \multicolumn{3}{c}{$\mathcal A=\mathcal T\tau$}\\
    \cmidrule(lr){3-5}\cmidrule(l){6-8}
    Order & Matrix part & $(\pi_c,\pi_{s_z})$ & Charge & Spin
      & $(\pi_c,\pi_{s_z})$ & Charge & Spin\\
    \midrule
    1 & $\operatorname{Re}X$ & $(+,-)$ & $\checkmark$ & $\usym{2613}$ & $(+,-)$ & $\checkmark$ & $\usym{2613}$\\
    1 & $\operatorname{Im}X$ & $(+,-)$ & $\checkmark$ & $\usym{2613}$ & $(-,+)$ & $\usym{2613}$ & $\checkmark$\\
    2 & $\operatorname{Re}X$ & $(-,+)$ & $\usym{2613}$ & $\checkmark$ & $(-,+)$ & $\usym{2613}$ & $\checkmark$\\
    2 & $\operatorname{Im}X$ & $(-,+)$ & $\usym{2613}$ & $\checkmark$ & $(+,-)$ & $\checkmark$ & $\usym{2613}$\\
    \bottomrule
  \end{tabular}
\end{table}

\subsection*{Spin-conserving clean dc limit and the six mechanisms}
We first combine all response terms at finite $\gamma$ before taking the dc limit. With conserved $\hat s_z$ and resolved interband gaps, $|\varepsilon_n-\varepsilon_m|\tau_r/\hbar\gg1$, the leading clean-limit decomposition gives Table~\ref{tab:mechanisms-comparison}. Under $\Theta$, forbidden terms vanish within each spin block; allowed components depend further on the interspin operation $O_s=[C_2^\perp\Vert O]$ and the full magnetic group. The six mechanism formulas, their symmetry transformations, and the component constraints are derived in SM Sec.~S2.2~\cite{supplemental}. When spin is not conserved, the applicable constraints are the full-vertex rules in Table~\ref{tab:full-response-selection} rather than assuming the same scalar mechanism decomposition.

\begin{table}[!htbp]
  \caption{\label{tab:mechanisms-comparison}
  Six mechanisms in the spin-conserving clean dc limit under the three symmetries,
  imposed separately. For the real dc kernels including occupation weights,
  $K_{g\mathcal N}^{X,\alpha}(-\bk)=\pi_{g,\alpha}^{X}K_{\mathcal N}^{X,\alpha}(\bk)$,
  where $X$ labels the mechanism and $\mathcal N$ the band indices.
  Odd kernels cancel between partners; $+$ and $-$ denote $\pm1$, with
  $\times$ and $\checkmark$ defined as in Table~\ref{tab:full-response-selection}.
  All columns assume common equilibrium occupations and equal constant $\tau_r$
  for symmetry partners. Powers of $\tau_r$ give the leading clean dc scaling;
  QMD denotes quantum metric dipole. Additional magnetic-group constraints and
  tensor identities can further eliminate allowed components.}
  \centering
  \small
  \renewcommand{\arraystretch}{1.22}
  \setlength{\tabcolsep}{4pt}
  \begin{tabular}{@{}lcc*{3}{ccc}@{}}
    \toprule
    & & & \multicolumn{3}{c}{Odd parity: $\Uop=[C_2^\perp\Vert P]$}
      & \multicolumn{3}{c}{Odd parity: $\mathcal T\tau$}
      & \multicolumn{3}{c}{Altermagnets: $\Theta=[C_2^\perp\mathcal T\Vert E]$}\\
    \cmidrule(lr){4-6}\cmidrule(lr){7-9}\cmidrule(l){10-12}
    Mechanism & $N$ & Scaling
      & $(\pi_c,\pi_{s_z})$ & Charge & Spin
      & $(\pi_c,\pi_{s_z})$ & Charge & Spin
      & $(\pi_c,\pi_{s_z})$ & Charge & Spin\\
    \midrule
    Drude & 1 & $\tau_r$
      & $(+,-)$ & $\checkmark$ & $\usym{2613}$
      & $(+,-)$ & $\checkmark$ & $\usym{2613}$
      & $(+,+)$ & $\checkmark$ & $\checkmark$\\
    Berry curvature (BC) & 1 & $\tau_r^0$
      & $(+,-)$ & $\checkmark$ & $\usym{2613}$
      & $(-,+)$ & $\usym{2613}$ & $\checkmark$
      & $(-,-)$ & $\usym{2613}$ & $\usym{2613}$\\
    Nonlinear Drude & 2 & $\tau_r^2$
      & $(-,+)$ & $\usym{2613}$ & $\checkmark$
      & $(-,+)$ & $\usym{2613}$ & $\checkmark$
      & $(-,-)$ & $\usym{2613}$ & $\usym{2613}$\\
    Berry curvature dipole & 2 & $\tau_r$
      & $(-,+)$ & $\usym{2613}$ & $\checkmark$
      & $(+,-)$ & $\checkmark$ & $\usym{2613}$
      & $(+,+)$ & $\checkmark$ & $\checkmark$\\
    Interband QMD & 2 & $\tau_r^0$
      & $(-,+)$ & $\usym{2613}$ & $\checkmark$
      & $(-,+)$ & $\usym{2613}$ & $\checkmark$
      & $(-,-)$ & $\usym{2613}$ & $\usym{2613}$\\
    Intraband QMD & 2 & $\tau_r^0$
      & $(-,+)$ & $\usym{2613}$ & $\checkmark$
      & $(-,+)$ & $\usym{2613}$ & $\checkmark$
      & $(-,-)$ & $\usym{2613}$ & $\usym{2613}$\\
    \bottomrule
  \end{tabular}
\end{table}

\end{document}


\title{Supplemental Material for\texorpdfstring{\\}{ }
General Transport Signatures of Odd-Parity Magnets: Charge--Spin Selection Rules and \texorpdfstring{Temporal--Spectral Separation}{Temporal--Spectral Separation}}
\author{Junxi Yu}
\author{Pu Zhang}
\author{Yichen Liu}
\author{Yongpan Li}
\author{Cheng-Cheng Liu}
\email{ccliu@bit.edu.cn}
\affiliation{Centre for Quantum Physics, Key Laboratory of Advanced Optoelectronic Quantum Architecture and Measurement (MOE), School of Physics, Beijing Institute of Technology, Beijing 100081, China}
\maketitle

\tableofcontents
\clearpage

\section{Driven-state covariance and current selection rules in collinear and noncollinear magnets}
\label{sec:covariance}

We derive the charge--spin current relations and harmonic selection rules for collinear and noncollinear magnets whose zero-field Hamiltonian preserves the unitary spin-flip inversion symmetry $\Uop=[C_2^\perp\Vert P]$.
The rotation axis of $C_2^\perp$ is perpendicular to the measured spin direction $z$.
The proof combines the covariance of the driven dynamics with the uniqueness of the physical steady state and applies to systems with conserved or nonconserved spin.
Throughout, the applied electric field is spatially uniform within the electric-dipole approximation.
Hats denote operators and their full matrix representations.

\subsection{Transformations of the Hamiltonian and current operators}
\label{sec:H-proof}

Taking the inversion center as the origin and choosing $x$ as the spin-rotation axis, we write
\begin{equation}
  \Uop=[C_{2x}^{\mathrm{spin}}||P],
  \qquad \Uop \hat H_0\Uop^{-1}=\hat H_0.
  \label{eq:U-definition}
\end{equation}
Spatial inversion reverses position, while the spin rotation reverses the measured $z$ component, giving
\begin{equation}
  \Uop\hat{\bm r}\Uop^{-1}=-\hat{\bm r},
  \qquad \Uop\hat s_z\Uop^{-1}=-\hat s_z,
  \qquad \Uop t\Uop^{-1}=t,
  \qquad \Uop i\Uop^{-1}=i.
  \label{eq:basic-transformations}
\end{equation}
The last equality expresses the unitary character of $\Uop$. 
It maps a band state $|\psi_{\nu\bk}\rangle$ to an equal-energy partner at $-\bk$, so that
\begin{equation}
  \varepsilon_{\nu}(-\bk)=\varepsilon_{\Uop\nu}(\bk),
  \qquad
  S_{z,\nu}(-\bk)=-S_{z,\Uop\nu}(\bk),
  \quad S_{z,\nu}(\bk)=\langle u_{\nu\bk}|\hat s_z|u_{\nu\bk}\rangle.
  \label{eq:band-pair}
\end{equation}
Here $\nu$ is the band index.
In the collinear, spin-conserving limit, this symmetry relation reduces to an exchange of the spin-up and spin-down bands.

In the length gauge, the driven Hamiltonian is
\begin{equation}
  \hat H[\bE]=\hat H_0-\hat{\bm d}\cdot\bE,
  \qquad \hat{\bm d}=-e\sum_a\hat{\bm r}_a,
  \label{eq:length-gauge}
\end{equation}
where the electron charge is $-e$ and $a$ labels electrons.
The electric dipole changes sign under $\Uop$, yielding
\begin{equation}
  \Uop \hat H[\bE]\Uop^{-1}
  =\hat H_0+\hat{\bm d}\cdot\bE=\hat H[-\bE].
  \label{eq:H-covariance}
\end{equation}
This equation relates the two driving configurations $+\bE$ and $-\bE$ and holds at every instant for a time-dependent drive.

In the full real-space Hilbert space, the electric-dipole coupling in the length gauge commutes with position, so the single-electron velocity can be written as
\begin{equation}
  \hat v_i=\frac{i}{\hbar}[\hat H[\bE],\hat r_i]
  =\frac{i}{\hbar}[\hat H_0,\hat r_i],
  \qquad \Uop\hat v_i\Uop^{-1}=-\hat v_i.
  \label{eq:velocity-parity}
\end{equation}
Expressing both currents in charge-equivalent units, as in the main text, we define
\begin{equation}
  \hat J_{c,i}=-e\hat v_i,
  \qquad
  \hat J_{s_z,i}=-\frac{e}{\hbar}\{\hat v_i,\hat s_z\}.
  \label{eq:current-definitions}
\end{equation}
The spin current is the conventional angular-momentum current $\{\hat v_i,\hat s_z\}/2$ multiplied by $-2e/\hbar$.
Since both velocity and $\hat s_z$ change sign, we obtain
\begin{equation}
  \Uop\hat J_{\alpha,i}\Uop^{-1}
  =\eta_\alpha\hat J_{\alpha,i},
  \qquad \eta_c=-1,\quad\eta_{s_z}=+1.
  \label{eq:inverse-parity}
\end{equation}

These parities follow from the action of $\Uop$ on the electric-field coupling and current operators.

\subsection{Covariance of the full dynamics and the physical density operator}
\label{sec:rho-proof}

We use a Liouville-space formulation, in which a Liouvillian superoperator generates the evolution of the density operator and includes both coherent and dissipative dynamics~\cite{bucaSymmetryLindblad2012}.
Define the conjugation map $\Uscr[\hat X]=\Uop \hat X\Uop^{-1}$ acting on operators, and write the dynamics, including scattering or coupling to a bath, as
\begin{equation}
  \partial_t\hat\rho(t)=\Lop_{\bE(t)}[\hat\rho(t)],
  \qquad
  \Lop_{\bE}[\hat X]=-
  \frac{i}{\hbar}[\hat H[\bE],\hat X]+\Iop_{\bE}[\hat X],
  \label{eq:Liouvillian}
\end{equation}
where $\Iop_{\bE}$ is the dissipative or collision term.
Equation~\eqref{eq:H-covariance} ensures covariance of the coherent part; the dissipative part must additionally satisfy
\begin{equation}
  \Uscr\,\Iop_{\bE}=\Iop_{-\bE}\,\Uscr.
  \label{eq:dissipative-covariance}
\end{equation}
This requires equal rates for symmetry-related scattering processes~\cite{marcianiBoltzmannTransport2021}, together with compatible bath, injection, and boundary conditions.
For example, a collision term describing relaxation toward a $\Uop$-invariant equilibrium state satisfies this condition if the same relaxation time is assigned to symmetry-related states.

Combining the coherent and dissipative parts gives the covariance of the full generator:
\begin{equation}
  \Uscr\,\Lop_{\bE}
  =\Lop_{-\bE}\,\Uscr.
  \label{eq:L-covariance}
\end{equation}
Let $\hat\rho_{\bE}(t)$ be a solution for a given field history $\bE(t)$.
Its partner $\widetilde{\hat\rho}(t)=\Uscr[\hat\rho_{\bE}(t)]$ then satisfies
\begin{equation}
  \partial_t\widetilde{\hat\rho}(t)
  =\Uscr\, [\partial_{t}\hat\rho_{\bE}(t)]
  =\Uscr\,\Lop_{\bE(t)}[\hat\rho_{\bE}(t)]
  =\Lop_{-\bE(t)}[\widetilde{\hat\rho}(t)].
  \label{eq:trajectory-proof}
\end{equation}
If the initial states are related by the same operation,
$\hat\rho_{-\bE}(t_0)=\Uscr[\hat\rho_{\bE}(t_0)]$,
uniqueness of the solution to the initial-value problem implies
\begin{equation}
  \hat\rho_{-\bE}(t)=\Uop\hat\rho_{\bE}(t)\Uop^{-1}.
  \label{eq:trajectory-covariance}
\end{equation}
This is the density-matrix relation used in the main text.
A common choice of shared initial state is the $\Uop$-invariant zero-field equilibrium state
\begin{equation}
  \hat\rho_0=Z_0^{-1}e^{-\beta(\hat H_0-\mu\hat N)},
  \qquad \Uscr[\hat\rho_0]=\hat\rho_0,
  \label{eq:equilibrium-density}
\end{equation}
where $\hat N$ is the particle-number operator.
Applying $\bE(t)$ and $-\bE(t)$ separately to this state produces two causal evolution trajectories satisfying Eq.~\eqref{eq:trajectory-covariance}.

Specifically, for a static electric field, if $\hat\rho_{\bE}$ is a steady state, Eq.~\eqref{eq:L-covariance} gives
\begin{equation}
  \Lop_{-\bE}[\Uscr[\hat\rho_{\bE}]]
  =\Uscr[\Lop_{\bE}[\hat\rho_{\bE}]]
  =\Uscr[\partial_{t}\hat\rho_{\bE}]
  =0.
  \label{eq:zero-mode-proof}
\end{equation}
Unitary conjugation maps $\hat\rho_{\bE}$ to a normalized physical steady state under the reversed field.
Uniqueness of this steady state for the specified preparation and compatible boundary conditions then gives
\begin{equation}
  \Uop\hat\rho_{\bE}\Uop^{-1}=\hat\rho_{-\bE}.
  \label{eq:rho-covariance}
\end{equation}
For finite-dimensional quantum Markov dynamics, sufficient conditions are known under which all initial states relax to the same steady state~\cite{spohnAlgebraicCondition1977}.

\subsection{Central current relations and all-order selection rules}
\label{sec:current-proof}

Combining Eqs.~\eqref{eq:inverse-parity} and~\eqref{eq:rho-covariance} with cyclicity of the trace yields
\begin{equation}
  \begin{aligned}
  J_{\alpha,i}(-\bE)
  &=\Tr\!\left(\Uop\hat\rho_{\bE}\Uop^{-1}\hat J_{\alpha,i}\right)\\
  &=\Tr\!\left(\hat\rho_{\bE}\Uop^{-1}\hat J_{\alpha,i}\Uop\right)
   =\eta_\alpha J_{\alpha,i}(\bE).
  \end{aligned}
  \label{eq:trace-proof}
\end{equation}
Here we used
$\Uop^{-1}\hat J_{\alpha,i}\Uop=\eta_\alpha\hat J_{\alpha,i}$,
which follows directly from Eq.~\eqref{eq:inverse-parity}. Every spatial component therefore obeys the two central relations of the main text,
\begin{equation}
  \bm J_c(\bE)=-\bm J_c(-\bE),
  \qquad \bm J_{s_z}(\bE)=+\bm J_{s_z}(-\bE).
  \label{eq:master-relations}
\end{equation}
These relations apply directly to the full currents and do not require a weak-field expansion.
For time-dependent driving, the same trace argument gives $\bm J_c[-\bE](t)=-\bm J_c[\bE](t)$
and $\bm J_{s_z}[-\bE](t)=\bm J_{s_z}[\bE](t)$.

The above symmetry forces the zero-field charge current to vanish, whereas the spin current may have an equilibrium background.
As in the main text, we henceforth use $J_\alpha$ to denote the field-induced current after subtraction of this background;
this subtraction preserves Eq.~\eqref{eq:master-relations}.
If the response admits an expansion around zero field, then
\begin{equation}
  J_{\alpha,i}(\bE)=\sum_{N\geq1}
  \chi_{ia_1\cdots a_N}^{\alpha,(N)}
  E_{a_1}\cdots E_{a_N},
  \qquad
  \begin{cases}
    \chi_c^{(2m)}=0,&m\geq1,\\
    \chi_{s_z}^{(2m+1)}=0,&m\geq0.
  \end{cases}
  \label{eq:order-selection}
\end{equation}

In the collinear, spin-conserving limit, $J_{s_z}=J_\uparrow-J_\downarrow$, where $J_\uparrow$ and $J_\downarrow$ are the charge currents in the two spin blocks.
For $\Uop=[C_{2x}\Vert P]$, the $y$- and $z$-polarized spin currents are even under electric-field reversal, while the $x$-polarized spin current is odd.

\subsection{Half-period dynamical symmetry and nonperturbative harmonic separation}
\label{sec:half-period-proof}
\label{sec:harmonic-proof}

Consider an electric field with period $T$ satisfying
\begin{equation}
  \bE(t+T)=\bE(t),
  \qquad \bE(t+T/2)=-\bE(t).
  \label{eq:half-wave-drive}
\end{equation}
Monochromatic drives with linear, elliptical, or circular polarization satisfy this condition,
as do periodic fields containing only odd input harmonics.
When the dissipative dynamics and boundary conditions, including any modulation of the baths or contacts, share the same half-period symmetry, the full generator obeys
\begin{equation}
  \Uscr\,\Lop(t)=\Lop(t+T/2)\,\Uscr,
  \qquad \Lop(t+T)=\Lop(t).
  \label{eq:half-period-L}
\end{equation}

Let $\hat\rho(t)$ be the $T$-periodic physical steady state reached after transients have decayed, and construct
$\widetilde{\hat\rho}(t)=\Uscr[\hat\rho(t+T/2)]$.
Equation~\eqref{eq:half-period-L} and periodicity give
\begin{equation}
  \partial_t\widetilde{\hat\rho}(t)
  =\Uscr\,\Lop(t+T/2)[\hat\rho(t+T/2)]=\Lop(t+T)\Uscr\,[\hat\rho(t+T/2)]
  =\Lop(t)[\widetilde{\hat\rho}(t)].
  \label{eq:half-period-partner-evolution}
\end{equation}
The transformed state is periodic, $\widetilde{\hat\rho}(t+T)=\widetilde{\hat\rho}(t)$, and remains a normalized physical density matrix.
At a fixed drive phase, uniqueness of the periodic steady state for compatible preparation conditions gives $\widetilde{\hat\rho}(t)=\hat\rho(t)$, or equivalently,
\begin{equation}
  \Uop\hat\rho(t+T/2)\Uop^{-1}=\hat\rho(t).
  \label{eq:half-period-rho-forward}
\end{equation}
This is the constraint imposed on the driven state by the half-period dynamical symmetry $(\Uop,T/2)$.

Substituting Eq.~\eqref{eq:half-period-rho-forward} into the current expectation values and again using Eq.~\eqref{eq:inverse-parity}, we find
\begin{equation}
  J_{c,i}(t+T/2)=-J_{c,i}(t),
  \qquad J_{s_z,i}(t+T/2)=+J_{s_z,i}(t).
  \label{eq:half-period-currents}
\end{equation}
With $\Omega=2\pi/T$, we use the Fourier convention of the main text,
\begin{equation}
  J_{\alpha,i}(t)=\sum_{n\in\mathbb Z}
    J_{\alpha,i}^{n\Omega}e^{-in\Omega t},
  \qquad
  J_{\alpha,i}^{n\Omega}
    =\frac{1}{T}\int_0^Tdt\,J_{\alpha,i}(t)e^{in\Omega t}.
  \label{eq:full-current-fourier}
\end{equation}
A half-period time translation multiplies the $n$th Fourier component by $(-1)^n$, so
\begin{equation}
  [(-1)^n-\eta_\alpha]J_{\alpha,i}^{n\Omega}=0,
  \qquad
  J_{c,i}^{2m\Omega}=0,\quad
  J_{s_z,i}^{(2m+1)\Omega}=0.
  \label{eq:complete-harmonic-selection}
\end{equation}
For nonnegative frequencies, $m\geq0$; the negative-frequency components follow from the complex-conjugation relation for real currents.
Only $\Omega,3\Omega,\ldots$ are allowed in the charge channel,
and only $0,2\Omega,4\Omega,\ldots$ in the spin channel.
This constraint applies to the full complex Fourier coefficients.
In particular, dc charge rectification vanishes, whereas field-induced spin rectification is allowed.

The response order $N$ counts powers of the field amplitude $E_0$, whereas the harmonic index $n$ labels the output frequency $n\Omega$.
For a monochromatic drive and time-translation-invariant zero-field dynamics, expanding in powers of $E_0$ gives
\begin{equation}
  J_{\alpha}^{n\Omega}=\sum_{l=0}^{\infty}C_{\alpha,n+2l}^{n\Omega}E_0^{n+2l}.
  \label{eq:harmonic-expand}
\end{equation}
For example, the first and second harmonics are
\begin{equation}
  \begin{aligned}
    J_c^\Omega&=C_{c,1}^{\Omega}E_0
      +C_{c,3}^{\Omega}E_0^3+\cdots,\\
    J_{s_z}^{2\Omega}&=C_{s_z,2}^{2\Omega}E_0^2+C_{s_z,4}^{2\Omega}E_0^4+\cdots.
  \end{aligned}
  \label{eq:harmonic-order-mixing}
\end{equation}
The symmetry of the periodic steady state enforces harmonic separation beyond the weak-field regime in both collinear and noncollinear systems.

\section{All transport mechanisms up to the second order under $\Uop=[C_2^\perp\Vert P]$, $\mathcal A=\mathcal T\tau$, and $\Theta=[C_2^\perp\mathcal T\Vert E]$}
\label{sec:microscopic-mechanisms}
\label{sec:ttau-response}
Nonlinear transport and related optical and magnetoelectric responses have drawn sustained interest as probes of band geometry, magnetic order, and dissipation~\cite{ulrichQuantumGeometric2026,shibataSecondOrderKeldysh2026,GuoDissipationGeometry2026,ParkerDiagrammatic2019,jiaNonlinearMagnetoelectric2026,zhangNonlinearChargeSpin2021,zhu_magnetic_2025}. This section derives the symmetry constraints on all transport mechanisms up to the second order listed in the End Matter of the main text:
linear Drude, Berry curvature (BC), nonlinear Drude (ND), Berry curvature dipole (BCD),
and interband and intraband quantum metric dipoles (inter-QMD and intra-QMD).
We consider the three symmetries $\Uop=[C_2^\perp\Vert P]$, $\mathcal A=\mathcal T\tau$, and $\Theta=[C_2^\perp\mathcal T\Vert E]$ separately.
Here $E$ denotes the spatial identity, and the rotation axis of $C_2^\perp$ is perpendicular to the measured spin direction $z$; $\tau$ is a fractional translation of the magnetic structure, distinct from the relaxation time $\tau_r$.

The comparison concerns spatially uniform electric fields and unit-cell-averaged bulk currents,
with a common equilibrium Fermi distribution and the same constant relaxation time $\tau_r$ for symmetry-related states.
The formulas below describe the clean dc decomposition within the independent-quasiparticle, constant-relaxation-time approximation
\cite{sundaramWavePacket1999,sodemannBerryCurvatureDipole2015,gaoFieldInducedPositionalShift2014,ulrichQuantumGeometric2026,shibataSecondOrderKeldysh2026}.
We first derive the constraints in the noncollinear case using the full spin-current operator, and then analyze the individual transport mechanisms in the collinear, spin-conserving limit using band and quantum-geometric quantities.

\subsection{Noncollinear case: channel selection from the full first- and second-order currents}
\label{sec:micro-noncollinear}

When spin is not conserved, the response requires the full spin-current matrix.
We derive the charge--spin constraints by expanding both the current operator and the driven electronic state up to second order.
Each contribution separates into a band-matrix product and a scalar propagation weight.
We track their transformations under $\mathcal U$ and $\mathcal A$ separately.

We use Peierls coupling and a constant scalar broadening $\gamma=\hbar/(2\tau_r)>0$ that is independent of the applied field.
The derivation starts at finite frequency and finite $\gamma$ and retains the full spin-current matrix throughout the expansion.
For the charge-only derivation, see Ref.~\cite{shibataSecondOrderKeldysh2026}.

\subsubsection*{The full spin current and its field expansion}

Let $\alpha=c,s_z$ label the measured channel and define
\begin{equation}
  \hat w_i^c=\hat v_i,\qquad
  \hat w_i^{s_z}=\frac{1}{\hbar}\{\hat v_i,\hat s_z\},\qquad
  \hat J_{\alpha,i}=-e \hat w_i^\alpha.
  \label{eq:micro-full-vertices}
\end{equation}
The operators $\hat w_i^\alpha$ are full matrices in orbital and spin space.
We write their matrix elements as $w_{i,nm}^\alpha=\langle u_n|\hat w_i^\alpha|u_m\rangle$,
and similarly $v_{i,nm}=\langle u_n|\hat v_i|u_m\rangle$ and $(s_z)_{nm}=\langle u_n|\hat s_z|u_m\rangle$.
Inserting a complete set of band states gives
\begin{equation}
  \begin{aligned}
    w_{i,nm}^{s_z}
      &=\frac{1}{\hbar}\sum_l
        \bigl[v_{i,nl}(s_z)_{lm}+(s_z)_{nl}v_{i,lm}\bigr],\\
    w_{i,nn}^{s_z}
      &=\frac{2}{\hbar}
        \left[v_{i,nn}(s_z)_{nn}
          +\operatorname{Re}\sum_{l\ne n}v_{i,nl}(s_z)_{ln}\right].
  \end{aligned}
  \label{eq:micro-spin-vertex-band}
\end{equation}
The diagonal spin-current matrix element includes off-diagonal velocity and spin matrix elements.
In both channels, the driving vertices are velocities, whereas the measurement vertex is $\hat v_i$ for charge and $\hat w_i^{s_z}$ for spin.
This distinction must be retained in geometric reductions of the spin response.

Take $\bE=-\partial_t\bm A$ and $\hat H_{\bm A}(\bk)=\hat H_0(\bk+e\bm A/\hbar)$.
To expand the driving Hamiltonian and measured current together, define
\begin{equation}
  \hat v_j=\frac{1}{\hbar}D_j\hat H_0,\qquad
  \hat V_{jk}=\frac{1}{\hbar}D_j\hat v_k,\qquad
  \hat w_{i;j}^\alpha=\frac{1}{\hbar}D_j\hat w_i^\alpha,\qquad
  \hat w_{i;jk}^\alpha=\frac{1}{\hbar^2}D_jD_k\hat w_i^\alpha.
  \label{eq:micro-contact-vertices}
\end{equation}

$D_j$ differentiates the physical operator while removing the apparent change caused by a momentum-dependent basis.
In the fixed orbital--spin basis of the present Peierls model it is simply $\partial_j\equiv\partial/\partial k_j$.
To see why a correction is needed in a moving basis, let $B(\bk)$ be the unitary matrix whose columns form that complete basis,
and write the operator matrix as $\hat O=B^\dagger\hat O^{\rm fix}B$.
Differentiating this expression and using $B^\dagger B=\id$ gives
\begin{equation*}
  \partial_j\hat O
    =B^\dagger(\partial_j\hat O^{\rm fix})B
      +i[\mathcal A_j^{\rm b},\hat O],
  \qquad \mathcal A_j^{\rm b}=iB^\dagger\partial_jB.
\end{equation*}
The commutator is entirely due to the basis change.
Subtracting it defines the covariant derivative used in nonlinear-response theory~\cite{venturaGaugeCovariances2017,ParkerDiagrammatic2019}:
\begin{equation}
  D_j\hat O\equiv\partial_j\hat O-i[\mathcal A_j^{\rm b},\hat O]
    =B^\dagger(\partial_j\hat O^{\rm fix})B.
  \label{eq:micro-full-covariant-derivative}
\end{equation}
Here $\mathcal A_j^{\rm b}$ is the matrix-valued basis connection; in the band basis,
$(\mathcal A_j^{\rm b})_{nm}=i\langle u_n|\partial_j u_m\rangle$.
Under a further basis rotation $B\to BR(\bk)$, one has $\hat O\to R^\dagger\hat O R$
and $D_j\hat O\to R^\dagger(D_j\hat O)R$.
Thus the derivative vertices describe the same physical operators in any such basis, whereas an ordinary derivative of the band-basis matrix also differentiates the basis itself.
Since $\partial_j\hat s_z^{\rm fix}=0$, the spin operator satisfies $D_j\hat s_z=0$ in every such basis.

With summation over repeated spatial indices understood, the field expansion reads
\begin{equation}
  \begin{aligned}
    \hat H_{\bm A}&=\hat H_0+\hat H_1+\hat H_2+O(A^3),\\
    \hat H_1&=eA_j\hat v_j,\qquad
    \hat H_2=\frac{e^2}{2}A_jA_k\hat V_{jk},\\
    \hat J_{\alpha,i}[\bm A]
      &=-e\left[\hat w_i^\alpha+eA_j\hat w_{i;j}^\alpha
                  +\frac{e^2}{2}A_jA_k\hat w_{i;jk}^\alpha\right]+O(A^3).
  \end{aligned}
  \label{eq:micro-peierls-expansion}
\end{equation}
Terms containing $\hat V_{jk}$, $\hat w_{i;j}^\alpha$, or $\hat w_{i;jk}^\alpha$ are called contact terms:
they arise from the explicit field dependence of the Hamiltonian or the measured operator.
In this Peierls model with a complete basis, $\hat V_{jk}=\hat V_{kj}$ and $\hat w_{i;jk}^\alpha=\hat w_{i;kj}^\alpha$.
For the charge channel, Eq.~\eqref{eq:micro-peierls-expansion} reproduces the current expansion of the charge-only theory.
For the spin channel, the driving vertices remain charge-velocity vertices,
while the measurement vertex and all its derivatives are replaced by their full spin-current counterparts.

\subsubsection*{Expanding the electronic state and the measured operator}

Expand the single-particle density matrix as $\hat\rho_{\bm A}=\hat\rho^{(0)}+\hat\rho^{(1)}+\hat\rho^{(2)}+\cdots$,
where the superscript denotes the order in the vector potential.
From $J_{\alpha,i}=\intk\Tr(\hat J_{\alpha,i}[\bm A]\hat\rho_{\bm A})$, the first-order current is
\begin{equation}
  J_{\alpha,i}^{(1)}
    =-e\intk\Tr\left[
       \hat w_i^\alpha\hat\rho^{(1)}
       +eA_j\hat w_{i;j}^\alpha\hat\rho^{(0)}\right],
  \label{eq:micro-first-order-observable}
\end{equation}
and the second-order current is
\begin{equation}
  J_{\alpha,i}^{(2)}
    =-e\intk\Tr\left[
       \hat w_i^\alpha\hat\rho^{(2)}
       +eA_j\hat w_{i;j}^\alpha\hat\rho^{(1)}
       +\frac{e^2}{2}A_jA_k\hat w_{i;jk}^\alpha\hat\rho^{(0)}\right].
  \label{eq:micro-second-order-observable}
\end{equation}
The two first-order terms arise from changes in the electronic state and in the measured operator, respectively.
At second order, the three terms represent the second-order change in the state,
first-order changes in both the state and the operator, and the second-order change in the operator.

\begin{figure}[!t]
  \centering
  \includegraphics[width=0.98\linewidth]{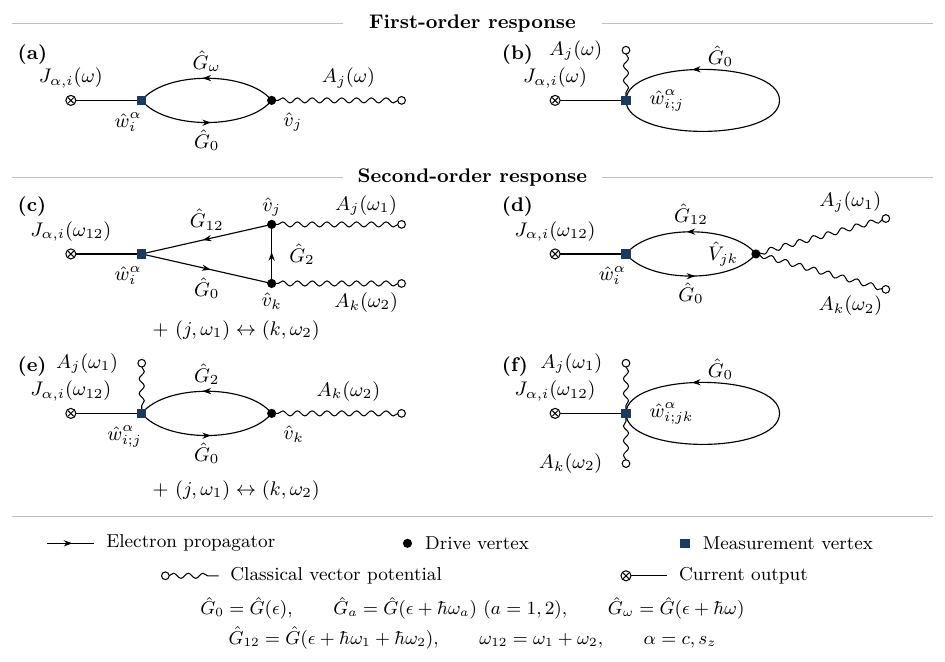}
  \caption{Field-vertex expansion of the full first- and second-order currents.
    Directed solid lines denote zero-field electronic Green functions, black dots denote driving vertices,
    blue squares denote measurement vertices, wavy lines denote the classical input vector potential,
    and a cross inside a circle marks the measured current.
    (a) First-order change in the electronic state; (b) first-order measurement contact term;
    (c) two first-order driving insertions; (d) one second-order driving insertion;
    (e) first-order changes in both the measurement operator and the electronic state;
    (f) second-order measurement contact term.
    Panels (c) and (e) also include the terms obtained by interchanging the two inputs,
    $(j,\omega_1)\leftrightarrow(k,\omega_2)$;
    the contact vertices in (d) and (f) are already symmetric under this interchange in the present model.
    Arrows follow the convention that $\hat G(t,t')$ points from $t'$ to $t$.
    The diagrams represent propagator chains before their Langreth decomposition;
    the full lesser projections and numerical coefficients are given in
    Eqs.~\eqref{eq:micro-general-first-order-keldysh} and \eqref{eq:micro-general-second-order-keldysh}.
    The charge and spin channels share the same driving vertices and differ at the measurement vertex $\hat w_i^\alpha$ and its derivatives,
    where $\alpha=c,s_z$.}
  \label{fig:micro-current-diagrams}
\end{figure}

With the time-domain normalization $G^<_{ab}(t,t')=(i/\hbar)\langle \hat c_b^\dagger(t')\hat c_a(t)\rangle$,
we have $\hat\rho^{(r)}(t)=-i\hbar \hat G^{(r)<}(t,t)$.
Let $\hat G$ denote the zero-field equilibrium Green operator (its full matrix in the chosen basis).
The Dyson equation is $\hat G_{\bm A}=\hat G+\hat G(\hat H_1+\hat H_2)\hat G_{\bm A}$,
where products at this stage denote time convolutions on the Keldysh contour.
Iterating and collecting powers of the field gives
\begin{equation}
  \hat G^{(1)}=\hat G\hat H_1\hat G,\qquad
  \hat G^{(2)}
    =\hat G\hat H_1\hat G\hat H_1\hat G
     +\hat G\hat H_2\hat G.
  \label{eq:micro-dyson-orders}
\end{equation}
The $\hat G^{(2)}$ in Eq.~\eqref{eq:micro-dyson-orders} describes two successive $\hat H_1$ insertions and one $\hat H_2$ insertion. Together with the field dependence of the measurement operator, they give the two first-order and four second-order contributions shown in Fig.~\ref{fig:micro-current-diagrams}.

We now express these propagator chains in frequency space and analyze the symmetry of the resulting matrix structures.

\subsubsection*{From the Dyson expansion to the full frequency response}

Using the $e^{-i\omega t}$ convention, $E_j(\omega)=i\omega A_j(\omega)$.
For equilibrium propagators, take
\begin{equation*}
  \hat G(t-t')=\int\frac{d\epsilon}{2\pi\hbar}\,
    e^{-i\epsilon(t-t')/\hbar}\hat G(\epsilon),
\end{equation*}
and write $\hat G_a=\hat G(\epsilon+\hbar\omega_a)$, $\hat G_0=\hat G(\epsilon)$, and $\omega_{12}=\omega_1+\omega_2$.
The energy origin is the same as in $\hat H_0$, and the chemical potential is included in $f_{\rm FD}$:
\begin{equation}
  \hat G^{R/A}(\epsilon)=[\epsilon-\hat H_0\pm i\gamma]^{-1},\qquad
  \hat G^<(\epsilon)=f_{\rm FD}(\epsilon)[\hat G^A(\epsilon)-\hat G^R(\epsilon)].
  \label{eq:micro-qmd-greens}
\end{equation}
After the Fourier transform, the factor of $\hbar$ in $\hat\rho=-i\hbar \hat G^<(t,t)$ cancels the factor in the energy-integration measure. Thus
\begin{equation}
  \hat\rho^{(0)}=-i\int\frac{d\epsilon}{2\pi}\hat G_0^<,\qquad
  \hat\rho^{(1)}(\omega)
    =-ieA_j(\omega)\int\frac{d\epsilon}{2\pi}
       [\hat G_\omega \hat v_j\hat G_0]^<.
  \label{eq:micro-rho-frequency}
\end{equation}
Substitution into Eq.~\eqref{eq:micro-first-order-observable} and division by $i\omega$ give the full linear electric-field response:
\begin{equation}
  \chi_{ij}^{\alpha,(1)}(\omega)
    =\frac{e^2}{\omega}\intk\int\frac{d\epsilon}{2\pi}
      \left\{\Tr[\hat w_i^\alpha \hat G_\omega \hat v_j\hat G_0]^<
                    +\Tr[\hat w_{i;j}^\alpha \hat G_0^<]\right\}.
  \label{eq:micro-general-first-order-keldysh}
\end{equation}
The two terms describe the first-order change in the electronic state and the first-order measurement contact term, shown in Figs.~\ref{fig:micro-current-diagrams}(a) and \ref{fig:micro-current-diagrams}(b), respectively.

The second-order derivation follows similar steps.
For $\hat G\hat H_1\hat G\hat H_1\hat G$, read the propagator chain from right to left:
the electron successively absorbs frequencies $\omega_2$ and $\omega_1$,
so its internal energies are $\epsilon$, $\epsilon+\hbar\omega_2$, and $\epsilon+\hbar\omega_{12}$.
This gives the chain $\hat G_{12}\hat v_j\hat G_2\hat v_k\hat G_0$.
For $\hat G\hat H_2\hat G$, a single $\hat H_2$ insertion carries both input frequencies and gives $\tfrac12\hat G_{12}\hat V_{jk}\hat G_0$.
Defining
\begin{equation}
  \int_{12}^{\omega_{\rm out}}
    \equiv\int\frac{d\omega_1}{2\pi}\frac{d\omega_2}{2\pi}
        2\pi\delta(\omega_{\rm out}-\omega_1-\omega_2),
  \label{eq:micro-frequency-convolution}
\end{equation}
we obtain
\begin{equation}
  \begin{aligned}
    \hat\rho^{(2)}(\omega_{\rm out})
      ={}&-ie^2\int_{12}^{\omega_{\rm out}}
         A_j(\omega_1)A_k(\omega_2)\int\frac{d\epsilon}{2\pi}\\
      &\times\left\{[\hat G_{12}\hat v_j\hat G_2\hat v_k\hat G_0]^<
                     +\frac12[\hat G_{12}\hat V_{jk}\hat G_0]^<\right\}.
  \end{aligned}
  \label{eq:micro-rho-second-frequency}
\end{equation}

One second-order term $\Tr[eA_j\hat w_{i;j}^{\alpha}\hat{\rho}^{(1)}]$ combines the linear field correction to the current operator with the linear change in the electronic state. Only frequency $\omega_2$ enters the electronic state, giving the matrix chain $\hat w_{i;j}^\alpha \hat G_2\hat v_k\hat G_0$; the other frequency, $\omega_1$, is carried by the explicit field dependence of the measurement operator. For the second-order measurement contact contribution, both inputs occur at the measurement operator and the electronic state remains in equilibrium, so the term contains only $\hat w_{i;jk}^\alpha \hat G_0^<$.

Substituting these results into Eq.~\eqref{eq:micro-second-order-observable}, we symmetrize over the input labels $(j,\omega_1)\leftrightarrow(k,\omega_2)$ while preserving the matrix order within each chain.
This leaves the classical-field convolution unchanged.
The two-insertion driving term and the first-order measurement contact term each acquire two input orderings; the other two contributions are already symmetric.
Finally, using $A_jA_k=-E_jE_k/(\omega_1\omega_2)$ gives
\begin{equation}
  \begin{aligned}
    \chi_{ijk}^{\alpha,(2)}(\omega_1,\omega_2)
      ={}&-\frac{ie^3}{2\omega_1\omega_2}
          \intk\int\frac{d\epsilon}{2\pi}
          \Big\{\Tr[\hat w_i^\alpha \hat G_{12}\hat v_j\hat G_2\hat v_k\hat G_0]^<\\
      &+\Tr[\hat w_i^\alpha \hat G_{12}\hat v_k\hat G_1\hat v_j\hat G_0]^<
        +\Tr[\hat w_i^\alpha \hat G_{12}\hat V_{jk}\hat G_0]^<\\
      &+\Tr[\hat w_{i;j}^\alpha \hat G_2\hat v_k\hat G_0]^<
        +\Tr[\hat w_{i;k}^\alpha \hat G_1\hat v_j\hat G_0]^<
        +\Tr[\hat w_{i;jk}^\alpha \hat G_0^<]\Big\}.
  \end{aligned}
  \label{eq:micro-general-second-order-keldysh}
\end{equation}
The response is defined by
$J_{\alpha,i}^{(2)}(\omega_{\rm out})=\int_{12}^{\omega_{\rm out}}\chi_{ijk}^{\alpha,(2)}E_j(\omega_1)E_k(\omega_2)$,
with no additional factor of $1/2$.
The first two terms are the two input orderings of Fig.~\ref{fig:micro-current-diagrams}(c);
the third corresponds to (d), the fourth and fifth are the two orderings of (e), and the sixth corresponds to (f).
These six traces comprise the four types of second-order contributions shown in Fig.~\ref{fig:micro-current-diagrams}.
To obtain the dc response, we first combine all terms at finite $\gamma$ and then send the input frequencies to zero, preserving the contact-term cancellations~\cite{shibataSecondOrderKeldysh2026}.

\subsubsection*{Spectral expansion: separating matrix structures from propagation weights}

The superscript $<$ denotes the lesser component of the entire ordered product.
For frequency-independent vertices $\hat O_1,\hat O_2$, the Langreth rules give~\cite{langrethResponseTheory1976,haugQuantumKinetics2008}
\begin{equation}
  \begin{aligned}
    {}[\hat G_a\hat O_1\hat G_b]^<
      &=\hat G_a^R\hat O_1\hat G_b^<+\hat G_a^<\hat O_1\hat G_b^A,\\
    [\hat G_a\hat O_1\hat G_b\hat O_2\hat G_c]^<
      &=\hat G_a^R\hat O_1\hat G_b^R\hat O_2\hat G_c^<
        +\hat G_a^R\hat O_1\hat G_b^<\hat O_2\hat G_c^A
        +\hat G_a^<\hat O_1\hat G_b^A\hat O_2\hat G_c^A.
  \end{aligned}
  \label{eq:micro-langreth-products}
\end{equation}
We next expand each propagator in band projectors.
Write $\hat H_0=\sum_n\varepsilon_n\hat P_n$, where $\hat P_n$ projects onto a band;
at a degeneracy, it projects onto the entire degenerate energy subspace.
Because the broadening is scalar,
\begin{equation}
  \begin{aligned}
    \hat G_a^{R/A/<}&=\sum_n\hat P_ng_{n,a}^{R/A/<},\\
    g_{n,a}^{R/A}
      &=\frac{1}{\epsilon+\hbar\omega_a-\varepsilon_n\pm i\gamma},\\
    g_{n,a}^<
      &=f_{\rm FD}(\epsilon+\hbar\omega_a)
          (g_{n,a}^A-g_{n,a}^R).
  \end{aligned}
  \label{eq:micro-band-propagators}
\end{equation}
Each propagator uses its own energy argument.
The projectors retain the eigenstate information, while the scalar functions $g$ contain the energy denominators, occupations, external frequencies, and broadening.

For the first term of Eq.~\eqref{eq:micro-general-second-order-keldysh}, for example, expand the three propagators and cyclically move the rightmost projector to the front of the trace:
\begin{equation}
  \begin{aligned}
    \Tr[\hat w_i^\alpha \hat G_{12}\hat v_j\hat G_2\hat v_k\hat G_0]^<
      &=\sum_{nml}
        \Tr(\hat P_n\hat w_i^\alpha \hat P_m\hat v_j\hat P_l\hat v_k)
        [g_{m,12}g_{l,2}g_{n,0}]^<,\\
    [g_{m,12}g_{l,2}g_{n,0}]^<
      &=g_{m,12}^Rg_{l,2}^Rg_{n,0}^<
        +g_{m,12}^Rg_{l,2}^<g_{n,0}^A
        +g_{m,12}^<g_{l,2}^Ag_{n,0}^A.
  \end{aligned}
  \label{eq:micro-spectral-factorization}
\end{equation}
All three Langreth components contain the same matrix product $Z$ (see Eq.~(\ref{eq:micro-contact-structures}) below); only their scalar propagation weights differ.
The remaining terms are treated in similar way.
At first order, only two closed matrix structures occur,
\begin{equation}
  L_{nm;ij}^\alpha=\Tr(\hat P_n\hat w_i^\alpha \hat P_m\hat v_j),\qquad
  C_{n;ij}^\alpha=\Tr[\hat P_n(D_j\hat w_i^\alpha)],
  \label{eq:micro-linear-structures}
\end{equation}
while at second order there are four,
\begin{equation}
  \begin{aligned}
    Z_{nml;ijk}^\alpha&=\Tr(\hat P_n\hat w_i^\alpha \hat P_m\hat v_j\hat P_l\hat v_k),\\
    Y_{nm;ijk}^\alpha&=\Tr[\hat P_n\hat w_i^\alpha \hat P_m(D_j\hat v_k)],\\
    W_{nm;ijk}^\alpha&=\Tr[\hat P_n(D_j\hat w_i^\alpha)\hat P_m\hat v_k],\\
    V_{n;ijk}^\alpha&=\Tr[\hat P_n(D_jD_k\hat w_i^\alpha)].
  \end{aligned}
  \label{eq:micro-contact-structures}
\end{equation}
Here $V_{n;ijk}^\alpha$ is a scalar trace, distinct from the coupling vertex $\hat V_{jk}$.
Figures~\ref{fig:micro-current-diagrams}(a) and \ref{fig:micro-current-diagrams}(b) generate $L$ and $C$, respectively,
while panels (c)--(f) generate $Z$, $Y$, $W$, and $V$.
The factors of $1/\hbar$ or $1/\hbar^2$ in the contact-vertex definitions are retained in the scalar propagation weights.
Table~\ref{tab:micro-all-structures} pairs each structure with its propagation weight,
showing explicitly that these six structures cover every term in
Eqs.~\eqref{eq:micro-general-first-order-keldysh} and \eqref{eq:micro-general-second-order-keldysh}.

\begin{table}[!htbp]
  \centering
  \caption{Spectral expansion of the full first- and second-order responses.
    Each matrix structure is multiplied by its scalar weight, then summed over band indices and integrated over energy and momentum.
    The common prefactors are $e^2/\omega$ at first order and $-ie^3/(2\omega_1\omega_2)$ at second order.
    The second-order $Z$ and $W$ contributions each also include a term with
    $(j,\omega_1)\leftrightarrow(k,\omega_2)$.
    Band indices may coincide; the expansion is not restricted to interband processes.}
  \label{tab:micro-all-structures}
  \small
  \renewcommand{\arraystretch}{1.38}
  \setlength{\tabcolsep}{7pt}
  \begin{tabular}{@{}clcc@{}}
    \toprule
    Order & Origin & Matrix structure & Scalar propagation weight\\
    \midrule
    1 & First-order change in the state & $L_{nm;ij}^\alpha$
      & $[g_{m,\omega}g_{n,0}]^<$\\
    1 & First-order measurement contact & $C_{n;ij}^\alpha$
      & $g_{n,0}^</\hbar$\\
    \midrule
    2 & Two first-order driving insertions & $Z_{nml;ijk}^\alpha$
      & $[g_{m,12}g_{l,2}g_{n,0}]^<$\\
    2 & One second-order driving insertion & $Y_{nm;ijk}^\alpha$
      & $[g_{m,12}g_{n,0}]^</\hbar$\\
    2 & First-order measurement contact & $W_{nm;ijk}^\alpha$
      & $[g_{m,2}g_{n,0}]^</\hbar$\\
    2 & Second-order measurement contact & $V_{n;ijk}^\alpha$
      & $g_{n,0}^</\hbar^2$\\
    \bottomrule
  \end{tabular}
\end{table}

``Closed'' means that the band indices of the matrix elements form a closed chain.
For nondegenerate bands, for example,
$Z_{nml;ijk}^\alpha=w_{i,nm}^\alpha v_{j,ml}v_{k,ln}$ follows the path $n\to m\to l\to n$,
so arbitrary Bloch-state phases cancel in the product.
The other structures share this phase invariance, and their projector form also ensures invariance under basis rotations within degenerate subspaces.

Both complete responses, for $N=1,2$, can therefore be written in the unified form
\begin{equation}
  \chi_\alpha^{(N)}
    =\sum_{a,\mathcal N}\intk
       {\mathcal W}_{\mathcal N}^{a,(N)}\,X_{\mathcal N}^{a,\alpha},
  \qquad
  \begin{cases}
    X^{a,\alpha}\in\{L^\alpha,C^\alpha\},&N=1,\\
    X^{a,\alpha}\in\{Z^\alpha,Y^\alpha,W^\alpha,V^\alpha\},&N=2.
  \end{cases}
  \label{eq:micro-full-spectral-response}
\end{equation}
Spatial indices, input frequencies, and required input permutations are suppressed here;
$\mathcal N$ collects the relevant band indices.
The coefficients $\mathcal W$ are obtained by integrating the scalar propagation weights in Table~\ref{tab:micro-all-structures} over energy and multiplying by the common prefactor.
For example, the coefficient of $Z_{nml;ijk}^\alpha$ is
$-ie^3/(2\omega_1\omega_2)\int d\epsilon/(2\pi)[g_{m,12}g_{l,2}g_{n,0}]^<$.
The symmetry argument requires only that this coefficient be equal for symmetry partners.

\subsubsection*{Tracking symmetry signs: a common factor at each response order}
\label{sec:micro-kernel-parity}

For both $g=\Uop$ and $g=\mathcal A$, velocity and $\hat s_z$ change sign,
so the measurement vertices have the same parities under the two operations:
$\eta_c=-1$ and $\eta_{s_z}=+1$.
Comparing matrix elements in symmetry-related band bases gives
\begin{equation}
  \begin{aligned}
    w_{i,g n\,g m}^\alpha(-\bk)
       &=\eta_\alpha\,\mathscr C_g[w_{i,nm}^\alpha(\bk)],\\
    v_{j,g n\,g m}(-\bk)
       &=-\mathscr C_g[v_{j,nm}(\bk)],\qquad
    \mathscr C_{\Uop}[z]=z,\quad
    \mathscr C_{\mathcal A}[z]=z^*.
  \end{aligned}
  \label{eq:micro-vertex-parity}
\end{equation}
State-dependent phases in a general gauge cancel in the closed structures,
as does the fractional-translation phase of $\mathcal A$.
Each additional physical momentum derivative $D_j$ contributes another minus sign.

Counting the signs in each of the six structures in Table~\ref{tab:micro-all-structures} gives
\begin{equation}
  \begin{array}{c|c|c}
    \text{Order and structure} & \text{Vertex signs} & \text{Total sign}\\ \hline
    N=1:\ L^\alpha & \eta_\alpha(-1) & -\eta_\alpha\\
    N=1:\ C^\alpha & (-1)\eta_\alpha & -\eta_\alpha\\ \hline
    N=2:\ Z^\alpha & \eta_\alpha(-1)(-1) & +\eta_\alpha\\
    N=2:\ Y^\alpha & \eta_\alpha[(-1)(-1)] & +\eta_\alpha\\
    N=2:\ W^\alpha & [(-1)\eta_\alpha](-1) & +\eta_\alpha\\
    N=2:\ V^\alpha & (-1)^2\eta_\alpha & +\eta_\alpha
  \end{array}
  \label{eq:micro-contact-parities}
\end{equation}
For $Y$, for example, one minus sign comes from $D_j$ and the other from $\hat v_k$.
For $W$, $D_j\hat w_i^\alpha$ contributes $-\eta_\alpha$, which is then multiplied by the minus sign from $\hat v_k$.
The contact terms therefore preserve the common sign at a given order:
whether a derivative acts at the driving or measurement end, the total sign is $q_\alpha^{(N)}=\eta_\alpha(-1)^N$.

Projectors only map band labels to their symmetry partners and introduce no additional signs.
For a closed trace, a unitary transformation does not complex conjugate the result,
whereas an antiunitary transformation complex conjugates the entire trace.
All six structures therefore obey
\begin{equation}
  X_{g\mathcal N}^{a,\alpha}(-\bk)
    =q_\alpha^{(N)}
      \begin{cases}
        X_{\mathcal N}^{a,\alpha}(\bk),&g=\Uop,\\
        [X_{\mathcal N}^{a,\alpha}(\bk)]^*,&g=\mathcal A,
      \end{cases}
  \label{eq:micro-all-structure-parity}
\end{equation}
This relation covers every structure in the table and all input permutations,
without assuming spin conservation or any symmetry under interchange of spatial indices.

We must also specify how the scalar weights transform.
Partner energies satisfy $\varepsilon_{g n}(-\bk)=\varepsilon_n(\bk)$, with the same occupations and broadening.
Inserting these values into Eq.~\eqref{eq:micro-band-propagators} at the same external frequencies and with the same retarded-response convention gives, for each propagator,
\begin{equation}
  g_{g n,a}^{R/A/<}(\epsilon,-\bk)
     =g_{n,a}^{R/A/<}(\epsilon,\bk),
  \qquad
  {\mathcal W}_{g\mathcal N}^{a,(N)}(-\bk)
     ={\mathcal W}_{\mathcal N}^{a,(N)}(\bk).
  \label{eq:micro-spectral-weights-pair}
\end{equation}
Equation~\eqref{eq:micro-spectral-weights-pair} compares response weights at symmetry-related momenta with the same retarded prescription.
Antiunitarily transforming the full propagator instead maps $\hat G^R$ to $\hat G^A$, as in Eq.~\eqref{eq:micro-antiunitary-causality}.
Thus, in the partner comparison relevant here, $\mathcal A$ complex conjugates the matrix products while the scalar weights on the same retarded branch remain equal.

\subsubsection*{Unitary symmetry: common channel selection for all first- and second-order contributions}

For the unitary symmetry $g=\Uop$,
Eqs.~\eqref{eq:micro-all-structure-parity} and \eqref{eq:micro-spectral-weights-pair}
give the sum of a structure and its partner as
\begin{equation}
  {\mathcal W}_{\mathcal N}X_{\mathcal N}^\alpha
    +{\mathcal W}_{\Uop\mathcal N}X_{\Uop\mathcal N}^\alpha
       =(1+q_\alpha^{(N)}){\mathcal W}_{\mathcal N}X_{\mathcal N}^\alpha.
  \label{eq:micro-unitary-pair-sum}
\end{equation}
The corresponding momenta $\bk$ and $-\bk$ are implicit, and the common labels $a,N$ are suppressed.
Since the Brillouin-zone integral and complete band sums are invariant under this relabeling,
any channel with $q_\alpha^{(N)}=-1$ must vanish.
This cancellation holds for both quadratures of the complex response at finite frequency.

At first order, $q_c^{(1)}=+1$ and $q_{s_z}^{(1)}=-1$,
so the spin contributions from $L$ and $C$ cancel separately between partners.
At second order, $q_c^{(2)}=-1$ and $q_{s_z}^{(2)}=+1$,
so the charge contributions from $Z,Y,W,V$ each cancel between partners.
The full response therefore satisfies
\begin{equation}
  \chi_{s_z,ij}^{(1)}(\omega)=0,\qquad
  \chi_{c,ijk}^{(2)}(\omega_1,\omega_2)=0.
  \label{eq:micro-noncollinear-unitary-result}
\end{equation}
The first-order charge and second-order spin responses are symmetry allowed under $\Uop$.
The cancellation holds separately for each contribution: the first-order spin terms in Fig.~\ref{fig:micro-current-diagrams}(a),(b) and the second-order charge terms in panels (c)--(f) each vanish after partner summation.
These constraints persist in the dc and clean limits and reproduce the all-order selection rule of Sec.~\ref{sec:covariance} up to second order.

\subsubsection*{Antiunitary symmetry: real and imaginary matrix structures select opposite channels}

For the antiunitary symmetry $g=\mathcal A$, a structure and its partner instead give
\begin{equation}
  \begin{aligned}
    {\mathcal W}_{\mathcal N}X_{\mathcal N}^\alpha
      +{\mathcal W}_{\mathcal A\mathcal N}X_{\mathcal A\mathcal N}^\alpha
      &={\mathcal W}_{\mathcal N}
          \bigl(X_{\mathcal N}^\alpha+q_\alpha^{(N)}
                             [X_{\mathcal N}^\alpha]^*\bigr)\\
      &={\mathcal W}_{\mathcal N}\left[
          (1+q_\alpha^{(N)})\operatorname{Re}X_{\mathcal N}^\alpha
          +i(1-q_\alpha^{(N)})\operatorname{Im}X_{\mathcal N}^\alpha
                       \right].
  \end{aligned}
  \label{eq:micro-antiunitary-pair-sum}
\end{equation}
For $q=+1$, partner summation retains $\operatorname{Re}X$ and cancels $\operatorname{Im}X$; for $q=-1$, it retains $\operatorname{Im}X$ and cancels $\operatorname{Re}X$.
Here the real and imaginary parts refer to the closed matrix product $X$, whose scalar coefficient ${\mathcal W}$ is generally complex.
Since
$\operatorname{Re}({\mathcal W}X)=\operatorname{Re}{\mathcal W}\,\operatorname{Re}X-\operatorname{Im}{\mathcal W}\,\operatorname{Im}X$,
imaginary matrix structures can contribute to the real dc response through $\operatorname{Im}{\mathcal W}$.

Substituting $q_\alpha^{(N)}$ for each order and channel yields Table~\ref{tab:micro-noncollinear-selection}.
The covariant derivatives preserve Hermiticity, so $C_{n;ij}^\alpha$ and $V_{n;ijk}^\alpha$ are sums of real diagonal matrix elements within each energy subspace. Both therefore belong to the $\operatorname{Re}X$ sector.

\begin{table}[!htbp]
  \centering
  \caption{Channel constraints on the full first- and second-order responses in noncollinear systems.
    Each group gives the partner parities $(\pi_c,\pi_{s_z})$ of the matrix structures and the resulting charge and spin channels.
    $\operatorname{Re}X$ and $\operatorname{Im}X$ denote the real and imaginary parts of the closed matrix products in
    Eq.~\eqref{eq:micro-full-spectral-response}; each retains its full propagation weight.
    $\times$ and $\checkmark$ denote symmetry-forbidden and symmetry-allowed contributions, respectively.
    The two symmetries are imposed separately, without assuming spin conservation or decomposing the response into specific microscopic mechanisms.}
  \label{tab:micro-noncollinear-selection}
  \small
  \renewcommand{\arraystretch}{1.3}
  \setlength{\tabcolsep}{8pt}
  \begin{tabular}{@{}cc*{2}{ccc}@{}}
    \toprule
    & & \multicolumn{3}{c}{$\Uop$}
        & \multicolumn{3}{c}{$\mathcal A$}\\
    \cmidrule(lr){3-5}\cmidrule(l){6-8}
    Order & Matrix part & $(\pi_c,\pi_{s_z})$ & Charge & Spin
        & $(\pi_c,\pi_{s_z})$ & Charge & Spin\\
    \midrule
    1 & $\operatorname{Re}X$ & $(+,-)$ & $\checkmark$ & $\usym{2613}$
                            & $(+,-)$ & $\checkmark$ & $\usym{2613}$\\
    1 & $\operatorname{Im}X$ & $(+,-)$ & $\checkmark$ & $\usym{2613}$
                            & $(-,+)$ & $\usym{2613}$ & $\checkmark$\\
    2 & $\operatorname{Re}X$ & $(-,+)$ & $\usym{2613}$ & $\checkmark$
                            & $(-,+)$ & $\usym{2613}$ & $\checkmark$\\
    2 & $\operatorname{Im}X$ & $(-,+)$ & $\usym{2613}$ & $\checkmark$
                            & $(+,-)$ & $\checkmark$ & $\usym{2613}$\\
    \bottomrule
  \end{tabular}
\end{table}

For real matrix structures, $\mathcal A$ gives the same channel selection as $\Uop$: charge channel at first order and spin channel at second order.
Imaginary matrix structures follow the opposite selection, allowing spin channel at first order and charge channel at second order.
Contributions from both sectors can therefore produce charge and spin responses at the same order, so $\mathcal A$ alone does not enforce a common channel selection across all contributions.

In the collinear, clean dc limit, these rules reproduce the results of the following subsection:
Drude, ND, and both QMD contributions follow the real-structure selection, whereas BC and BCD follow the imaginary-structure selection.
For noncollinear systems, Eq.~\eqref{eq:micro-antiunitary-pair-sum} applies directly to the full matrix structures.

\subsection{Collinear case: from basic transformations to selection rules for all mechanisms}
\label{sec:micro-collinear}

The analysis above establishes the channel constraints for the full response without assuming spin conservation. To connect these general results to individual transport mechanisms, we now specialize to the clean dc limit of collinear, spin-conserving systems, where the responses can be expressed in terms of band dispersions and quantum-geometric quantities.

\subsubsection*{Spin blocks and the three symmetry operations}

For $[\hat H_0,\hat s_z]=0$, the Bloch states can be written as $|u_{ns}(\bk)\rangle$,
where $n$ is an orbital band index and $s=\uparrow,\downarrow$.
We denote the opposite spin by $\bar s$ and write the eigenvalue of $\hat s_z$ as $s_z=\xi_s\hbar/2$,
with $\xi_\uparrow=+1$ and $\xi_\downarrow=-1$.
Within each spin block, the spin current in the charge-equivalent units used in the main text satisfies
\begin{equation}
  \hat J_{c,i}\big|_s=-e\hat v_i\big|_s,\qquad
  \hat J_{s_z,i}\big|_s
    =-\frac{e}{\hbar}\{\hat v_i,\hat s_z\}\big|_s
    =\xi_s\hat J_{c,i}\big|_s.
  \label{eq:collinear-current-blocks}
\end{equation}
Thus, if $\chi_s^X$ denotes the charge-response coefficient of mechanism $X$ in a single spin block,
\begin{equation}
  \chi_c^X=\chi_\uparrow^X+\chi_\downarrow^X,\qquad
  \chi_{s_z}^X=\chi_\uparrow^X-\chi_\downarrow^X.
  \label{eq:micro-channel-sums}
\end{equation}
We suppress the spatial indices common to both sides of this relation.
Our conventions are $J_i^{(1)}=\chi_{ij}E_j$ for linear response and
$J_i^{(2)}=\chi_{ijk}E_jE_k$, with $\chi_{ijk}=\chi_{ikj}$, for second-order response;
no additional factor of $1/2!$ is included.

All three operations map $\bk$ to $-\bk$, but act differently on spin.
In $\Uop$, $C_2^\perp$ flips spin and $P$ reverses momentum; the combined operation is unitary.
In $\mathcal A$, time reversal flips both momentum and spin.
The translation leaves these quantities unchanged but contributes a phase to the Bloch state.
By contrast, $\Theta$ contains two spin flips, from $\mathcal T$ and $C_2^\perp$, which cancel and therefore preserve the spin block.
Writing $g n$ for the transformed orbital band label, we have
\begin{equation}
  \begin{aligned}
    \Uop &: (n,s,\bk)\longmapsto(\Uop n,\bar s,-\bk),\\
    \Theta &: (n,s,\bk)\longmapsto(\Theta n,s,-\bk),\\
    \mathcal A &: (n,s,\bk)\longmapsto(\mathcal A n,\bar s,-\bk).
  \end{aligned}
  \label{eq:collinear-state-pairing}
\end{equation}
A partner need not retain the original band index; all subsequent band sums include this relabeling.

\begin{table}[!htbp]
  \centering
  \caption{Action of the three symmetry operations on basic kinematic quantities.
    The electric-field row refers to a static uniform field; for a time-dependent field,
    $\Theta$ and $\mathcal A$ also map $t$ to $-t$.
    The two current rows describe the transformations of the operators themselves.}
  \label{tab:collinear-kinematics}
  \small
  \renewcommand{\arraystretch}{1.22}
  \begin{tabular}{@{}lccc@{}}
    \toprule
    Property or quantity &  $\Uop=[C_2^\perp\Vert P]$  &  $\Theta=[C_2^\perp\mathcal T\Vert E]$  &  $\mathcal A=\mathcal T\tau$\\
    \midrule
    Unitarity & Unitary & Antiunitary & Antiunitary\\
    Imaginary unit $i$ & $i$ & $-i$ & $-i$\\
    Time $t$ & $t$ & $-t$ & $-t$\\
    Momentum $\bk$ & $-\bk$ & $-\bk$ & $-\bk$\\
    Spin block $s$ & $\bar s$ & $s$ & $\bar s$\\
    Spin $\hat s_z$ & $-\hat s_z$ & $+\hat s_z$ & $-\hat s_z$\\
    Electric field $E_i$ & $-E_i$ & $+E_i$ & $+E_i$\\
    Velocity $\hat v_i$ & $-\hat v_i$ & $-\hat v_i$ & $-\hat v_i$\\
    Charge current $\hat J_{c,i}$ & $-\hat J_{c,i}$ & $-\hat J_{c,i}$ & $-\hat J_{c,i}$\\
    Spin current $\hat J_{s_z,i}$ & $+\hat J_{s_z,i}$ & $-\hat J_{s_z,i}$ & $+\hat J_{s_z,i}$\\
    \bottomrule
  \end{tabular}
\end{table}

Table~\ref{tab:collinear-kinematics} summarizes these kinematic transformations.
Antiunitary symmetries also transform the causal boundary conditions.
We therefore obtain the transport constraints by applying the equilibrium band and geometric transformations to the retarded response formulas.

\subsubsection*{Energies, occupations, and quantum-geometric quantities}

For a unified notation, let $s_{\Uop}=s_{\mathcal A}=\bar s$ and $s_\Theta=s$.
When $g$ is a symmetry of the zero-field Hamiltonian, partner states have equal energies:
\begin{equation}
  \varepsilon_{g n,s_g}(-\bk)=\varepsilon_{ns}(\bk),
  \qquad g\in\{\Uop,\Theta,\mathcal A\}.
  \label{eq:collinear-energy-pair}
\end{equation}
At a common chemical potential and temperature, define
$f_{ns}=f_{\rm FD}(\varepsilon_{ns})$,
$f'_{ns}=df_{\rm FD}/d\varepsilon|_{\varepsilon_{ns}}$, and
$\varepsilon_{nm,s}=\varepsilon_{ns}-\varepsilon_{ms}$.
The occupations $f$, their energy derivatives $f'$, and the ordered band-pair energy differences $\varepsilon_{nm,s}$ are unchanged between symmetry partners.

The signs of momentum derivatives follow from the chain rule.
If a real quantity obeys $X_{g n,s_g}(-\bk)=p_g^X X_{ns}(\bk)$, then
\begin{equation}
  [\partial_aX]_{g n,s_g}(-\bk)
    =-p_g^X[\partial_aX]_{ns}(\bk),
  \qquad
  p_g^{\partial_{a_1}\cdots\partial_{a_r}X}=(-1)^r p_g^X.
  \label{eq:collinear-derivative-rule}
\end{equation}
The brackets mean that the function is first differentiated with respect to its own momentum argument and then evaluated at $-\bk$.
The factor $p_g^X=\pm1$ is the sign acquired by $X$ under the partner mapping $\bk\to-\bk$ generated by $g$.
For example, $v_{i,ns}=\hbar^{-1}\partial_i\varepsilon_{ns}$ is odd,
the second derivative of the energy is even, and its third derivative is odd.

For geometric quantities, we must also distinguish unitary from antiunitary operations.
Within a spin block, define the band-pair quantum geometric tensor
\begin{equation}
  \mathcal Q_{ij,s}^{nm}
    =\langle\partial_i u_{ns}|u_{ms}\rangle
      \langle u_{ms}|\partial_j u_{ns}\rangle
    =\frac{\hbar^2v_{i,s}^{nm}v_{j,s}^{mn}}{\varepsilon_{nm,s}^2},
  \qquad m\ne n,
  \label{eq:collinear-qgt}
\end{equation}
and
\begin{equation}
  \begin{aligned}
    \Omega_{ij,s}^{nm}&=-2\operatorname{Im}\mathcal Q_{ij,s}^{nm},
    &\mathcal G_{ij,s}^{nm}&=\operatorname{Re}\mathcal Q_{ij,s}^{nm},\\
    \Omega_{ij,s}^{n}&=\sum_{m\ne n}\Omega_{ij,s}^{nm},
    &\mathcal G_{ij,s}^{n}&=\sum_{m\ne n}\mathcal G_{ij,s}^{nm},
    &\mathcal M_{ij,s}^{nm}&=\frac{\mathcal G_{ij,s}^{nm}}{\varepsilon_{nm,s}}.
  \end{aligned}
  \label{eq:micro-collinear-metric}
\end{equation}
Here $\Omega^n$ is the Berry curvature, $\mathcal G^n$ is the ordinary quantum metric,
and $\mathcal M^{nm}$ is the metric weighted by the inverse interband energy difference.
These quantities are defined using physical velocity matrix elements, so Bloch-state phases cancel in the band-pair products.
Under $\mathcal A$, the fractional-translation phase also cancels in these gauge-invariant quantities entering the bulk response.

Each velocity matrix element contributes a minus sign, leaving their product unchanged;
an antiunitary operation additionally complex conjugates the product. Hence
\begin{equation}
  \mathcal Q_{ij,s_g}^{g n\,g m}(-\bk)
    =\begin{cases}
      \mathcal Q_{ij,s}^{nm}(\bk),&g=\Uop,\\
      [\mathcal Q_{ij,s}^{nm}(\bk)]^*,&g=\Theta,\mathcal A.
    \end{cases}
  \label{eq:collinear-qgt-transform}
\end{equation}
Complex conjugation leaves the real part unchanged but reverses the imaginary part.
The quantum metric is therefore even under all three operations, whereas the Berry curvature is even under $\Uop$ and odd under $\Theta$ and $\mathcal A$.
Because $\varepsilon_{nm,s}$ is also even, $\mathcal M^{nm}$ transforms with the same sign as the ordinary metric.
Table~\ref{tab:collinear-basic-parities} summarizes these results and the signs of their momentum derivatives.

\begin{table}[!htbp]
  \centering
  \caption{For each quantity $X$, the sign $p_g[X]$ relates its value at $(-\bk,s_g)$ to that at $(\bk,s)$, with band indices mapped as $n\mapsto gn$ and $m\mapsto gm$. For $\mathcal U$ and $\mathcal A$, these signs relate opposite spin blocks at opposite momenta; for $\Theta$, they describe momentum parity within each spin block.}
  \label{tab:collinear-basic-parities}
  \small
  \renewcommand{\arraystretch}{1.27}
  \setlength{\tabcolsep}{10pt}
  \begin{tabular}{@{}lccc@{}}
    \toprule
    Quantity $X$ &  $\Uop=[C_2^\perp\Vert P]$: $s\to\bar s$
      & $\Theta=[C_2^\perp\mathcal T\Vert E]$: $s\to s$ & $\mathcal A=\mathcal T\tau$: $s\to\bar s$\\
    \midrule
    $\varepsilon_{ns}$, $\varepsilon_{nm,s}$ & $+$ & $+$ & $+$\\
    $f_{ns}$, $f'_{ns}$ & $+$ & $+$ & $+$\\
    $\partial_i\varepsilon_{ns}$, $v_{i,ns}$ & $-$ & $-$ & $-$\\
    $\partial_i\partial_j\varepsilon_{ns}$ & $+$ & $+$ & $+$\\
    $\partial_i\partial_j\partial_k\varepsilon_{ns}$ & $-$ & $-$ & $-$\\
    $\Omega_{ij,s}^{nm}$, $\Omega_{ij,s}^{n}$ & $+$ & $-$ & $-$\\
    $\mathcal G_{ij,s}^{nm}$, $\mathcal G_{ij,s}^{n}$ & $+$ & $+$ & $+$\\
    $\partial_\ell\Omega_{ij,s}^{n}$ & $-$ & $+$ & $+$\\
    $\partial_\ell\mathcal G_{ij,s}^{n}$ & $-$ & $-$ & $-$\\
    $\mathcal M_{ij,s}^{nm}=\mathcal G_{ij,s}^{nm}/\varepsilon_{nm,s}$ & $+$ & $+$ & $+$\\
    $\partial_\ell\mathcal M_{ij,s}^{nm}$ & $-$ & $-$ & $-$\\
    \bottomrule
  \end{tabular}
\end{table}

The single-band expressions containing energy-gap denominators apply in the clean limit with resolved interband separations,
$|\varepsilon_{nm,s}|\tau_r/\hbar\gg1$.
At degeneracies, one must use projectors onto symmetry-compatible subspaces, traces, or suitable expressions with finite broadening.
The symmetry pairing itself does not require nondegenerate band labels.

\subsubsection*{From single-spin integrands to charge- and spin-current kernels}

Write the single-spin response of mechanism $X$ as
$\chi_s^X=C_X\sum_{\mathcal N}\intk F_{\mathcal N,s}^X(\bk)$,
where $C_X$ collects factors independent of momentum and spin,
$F^X$ contains the occupations and band or geometric quantities, and $\mathcal N$ collects band indices.
Here $\intk$ denotes integration over the full Brillouin zone with measure $d^d k/(2\pi)^d$.
Multiplying the parities of its factors gives the transformation of each mechanism's integrand under $g$
\begin{equation}
  F_{g\mathcal N,s_g}^X(-\bk)
    =p_g^X F_{\mathcal N,s}^X(\bk).
  \label{eq:collinear-integrand-parity}
\end{equation}
Changing variables $\bk\to-\bk$ and relabeling the band sums gives
\begin{equation}
  \begin{array}{ll}
    g=\Uop,\mathcal A:&
      \chi_\downarrow^X=p_g^X\chi_\uparrow^X,\quad
      \chi_c^X=(1+p_g^X)\chi_\uparrow^X,\quad
      \chi_{s_z}^X=(1-p_g^X)\chi_\uparrow^X,\\[2pt]
    g=\Theta:&
      \chi_s^X=p_\Theta^X\chi_s^X,\quad s=\uparrow,\downarrow.
  \end{array}
  \label{eq:collinear-integrated-rule}
\end{equation}
This yields two distinct cancellation patterns.
For $\Uop$ and $\mathcal A$, $p_g^X=+1$ means that the two spin responses are equal:
their sum can give a charge response, while their difference, the spin response, vanishes.
The opposite holds for $p_g^X=-1$.
For $\Theta$, $p_\Theta^X=-1$ makes the integral vanish within each spin block, so both channels vanish;
$p_\Theta^X=+1$ alone imposes no relation between the two spin blocks.

The table in the main text uses the kernel parities $(\pi_c,\pi_{s_z})$, which include the measured current,
rather than the parity $p_g^X$ of the single-spin charge integrand.
In the collinear limit, the two kernels are proportional to $F_{\mathcal N,s}^X$ and $\xi_sF_{\mathcal N,s}^X$, respectively.
Since $\xi_{\bar s}=-\xi_s$, we obtain
\begin{equation}
  (\pi_{g,c}^X,\pi_{g,s_z}^X)
    =\begin{cases}
      (p_g^X,-p_g^X),&g=\Uop,\mathcal A,\\
      (p_g^X,p_g^X),&g=\Theta.
    \end{cases}
  \label{eq:collinear-current-kernel-map}
\end{equation}
A kernel with $\pi=-1$ cancels between symmetry partners, while $\pi=+1$ marks a symmetry-allowed channel.
For each mechanism below, we first determine $p_g^X$ from Table~\ref{tab:collinear-basic-parities},
then use Eqs.~\eqref{eq:collinear-integrated-rule} and \eqref{eq:collinear-current-kernel-map} to obtain the channel constraints.

\subsubsection*{Analysis of the six mechanisms}

\noindent\textbf{(1) Linear Drude: a product of two odd velocities.}\quad
The linear Drude response arises from the first-order change in the intraband occupation induced by the electric field.
For a single spin block,
\begin{equation}
  \chi_{ij,s}^{\mathrm D}
    =-\frac{e^2\tau_r}{\hbar^2}\sum_n\intk
       f'_{ns}(\partial_i\varepsilon_{ns})(\partial_j\varepsilon_{ns}).
  \label{eq:collinear-Drude}
\end{equation}
Under all three operations, $f'$ is even and both first derivatives of the energy are odd, so
\begin{equation}
  p_{\Uop}^{\mathrm D}
    =p_\Theta^{\mathrm D}
    =p_{\mathcal A}^{\mathrm D}
    =(+)(-)(-)=+.
  \label{eq:collinear-Drude-parity}
\end{equation}
Either $\Uop$ or $\mathcal A$ therefore gives $\chi_\uparrow^{\mathrm D}=\chi_\downarrow^{\mathrm D}$:
the charge Drude response is allowed and the spin Drude response cancels, corresponding to $(\pi_c,\pi_{s_z})=(+,-)$.
Under $\Theta$, the integrand is even within each spin block.
Both single-spin integrals are allowed and need not be equal, so both the charge and spin channels are allowed.

\medskip
\noindent\textbf{(2) Linear Berry curvature: complex conjugation changes the selection rule.}\quad
The BC contribution is
\begin{equation}
  \chi_{ij,s}^{\mathrm{BC}}
    =-\frac{e^2}{\hbar}\sum_n\intk f_{ns}\Omega_{ij,s}^{n}.
  \label{eq:collinear-BC}
\end{equation}
The occupation $f$ is unchanged, so this term inherits the partner parity of the Berry curvature:
\begin{equation}
  p_{\Uop}^{\mathrm{BC}}=(+)(+)=+,\qquad
  p_\Theta^{\mathrm{BC}}=p_{\mathcal A}^{\mathrm{BC}}=(+)(-)=-.
  \label{eq:collinear-BC-parity}
\end{equation}
Under $\Uop$, the two spin BC responses are equal: charge is allowed and spin vanishes, with kernel parities $(+,-)$.
Under $\mathcal A$, the two spin BC responses are opposite: charge vanishes and spin is allowed, with kernel parities $(-,+)$.
Under $\Theta$, $f_{ns}\Omega_{ij,s}^{n}$ is odd within each spin block,
so $\chi_\uparrow^{\mathrm{BC}}=\chi_\downarrow^{\mathrm{BC}}=0$.
Both channels are forbidden, with kernel parities $(-,-)$.
Thus, $\Uop$ and $\mathcal A$ select opposite channels for the linear BC response despite enforcing the same odd-parity band relation.

\medskip
\noindent\textbf{(3) Nonlinear Drude: the third energy derivative is odd.}\quad
The second-order intraband occupation produces the ND contribution.
Integration by parts over the Brillouin zone gives the leading clean-limit term
\begin{equation}
  \chi_{ijk,s}^{\mathrm{ND}}
    =-\frac{e^3\tau_r^2}{\hbar^3}\sum_n\intk
       f_{ns}\partial_i\partial_j\partial_k\varepsilon_{ns}.
  \label{eq:collinear-ND}
\end{equation}
This term is determined entirely by the band dispersion.
Since $f$ is even and the third derivative of the energy is odd,
\begin{equation}
  p_{\Uop}^{\mathrm{ND}}
    =p_\Theta^{\mathrm{ND}}
    =p_{\mathcal A}^{\mathrm{ND}}
    =(+)(-)=-.
  \label{eq:collinear-ND-parity}
\end{equation}
Either $\Uop$ or $\mathcal A$ gives $\chi_\downarrow^{\mathrm{ND}}=-\chi_\uparrow^{\mathrm{ND}}$:
charge cancels and spin is allowed, with kernel parities $(-,+)$.
Under $\Theta$, the ND integral vanishes separately in each spin block,
forbidding both channels, with kernel parities $(-,-)$.

\medskip
\noindent\textbf{(4) Berry curvature dipole: one additional momentum derivative.}\quad
The BCD response combines the Berry-geometric current with the first-order nonequilibrium occupation
\cite{sodemannBerryCurvatureDipole2015}. Symmetrizing the input indices yields
\begin{equation}
  \chi_{ijk,s}^{\mathrm{BCD}}
    =\frac{e^3\tau_r}{2\hbar^2}\sum_n\intk f_{ns}
       \bigl(\partial_k\Omega_{ij,s}^{n}
             +\partial_j\Omega_{ik,s}^{n}\bigr).
  \label{eq:collinear-BCD}
\end{equation}
Each summand contains one more momentum derivative than the BC term, reversing its parity:
\begin{equation}
  p_{\Uop}^{\mathrm{BCD}}=-,\qquad
  p_\Theta^{\mathrm{BCD}}=p_{\mathcal A}^{\mathrm{BCD}}=+.
  \label{eq:collinear-BCD-parity}
\end{equation}
Under $\Uop$, the two spin BCD responses are opposite: charge vanishes and spin is allowed, with kernel parities $(-,+)$.
Under $\mathcal A$, they are equal: charge is allowed and spin vanishes, with kernel parities $(+,-)$.
Under $\Theta$, the Berry curvature is odd within each spin block, but its momentum derivative is even.
Both BCD integrals are therefore allowed, and both current kernels are even, corresponding to $(+,+)$.
An antiunitary symmetry can thus forbid the linear BC response while allowing the second-order BCD response.

\medskip
\noindent\textbf{(5) Interband QMD: the energy-gap weight preserves the metric parity.}\quad
In the clean dc decomposition specified above, the interband QMD contribution is
\begin{equation}
  \chi_{ijk,s}^{\mathrm{inter\text{-}QMD}}
    =-\frac{e^3}{\hbar}\sum_n\sum_{m\ne n}\intk f_{ns}
      \bigl(2\partial_i\mathcal M_{jk,s}^{nm}
          -\partial_j\mathcal M_{ki,s}^{nm}
          -\partial_k\mathcal M_{ij,s}^{nm}\bigr).
  \label{eq:micro-inter-QMD}
\end{equation}
Each symmetry maps the ordered pair $(n,m)$ to $(gn,gm)$, preserving both $\mathcal G^{nm}$ and $\varepsilon_{nm}$. Their ratio $\mathcal M^{nm}$ is therefore even, and each momentum derivative in the interband QMD expression is odd. Thus
\begin{equation}
  p_{\Uop}^{\mathrm{inter\text{-}QMD}}
    =p_\Theta^{\mathrm{inter\text{-}QMD}}
    =p_{\mathcal A}^{\mathrm{inter\text{-}QMD}}
    =(+)(-)=-.
  \label{eq:collinear-inter-QMD-parity}
\end{equation}
Under either $\Uop$ or $\mathcal A$, the two spin responses are opposite,
so charge cancels and spin is allowed, with kernel parities $(-,+)$.
Under $\Theta$, each spin integral vanishes separately, forbidding both channels, with kernel parities $(-,-)$.

\medskip
\noindent\textbf{(6) Intraband QMD: the momentum derivative of the ordinary metric is odd.}\quad
The intraband QMD contribution is
\begin{equation}
  \chi_{ijk,s}^{\mathrm{intra\text{-}QMD}}
    =\frac{e^3}{2\hbar}\sum_n\intk
       f'_{ns}\partial_i\mathcal G_{jk,s}^{n}.
  \label{eq:micro-intra-QMD}
\end{equation}
Here $f'$ and $\mathcal G^n$ are even, whereas $\partial_i\mathcal G^n$ is odd. Hence
\begin{equation}
  p_{\Uop}^{\mathrm{intra\text{-}QMD}}
    =p_\Theta^{\mathrm{intra\text{-}QMD}}
    =p_{\mathcal A}^{\mathrm{intra\text{-}QMD}}
    =(+)(-)=-.
  \label{eq:collinear-intra-QMD-parity}
\end{equation}
This gives the same selection rule as for the interband QMD:
under either $\Uop$ or $\mathcal A$, charge vanishes and spin is allowed, with kernel parities $(-,+)$;
under $\Theta$, both channels vanish, with kernel parities $(-,-)$.

The two QMD contributions have the same parity but different microscopic weights.
The intra-QMD term is weighted by $f'$ and is confined to the Fermi surface at zero temperature.
The inter-QMD term contains an additional energy-gap weight.
Equation~\eqref{eq:micro-inter-QMD} expresses it as a Fermi-sea integral weighted by $f$;
integration by parts gives an equivalent Fermi-surface form
\cite{gaoFieldInducedPositionalShift2014,ulrichQuantumGeometric2026,shibataSecondOrderKeldysh2026}.
Both QMD terms scale as $\tau_r^0$ to leading order, whereas ND and BCD scale as $\tau_r^2$ and $\tau_r$, respectively.
These lifetime scalings affect the response magnitude but not the symmetry cancellations derived above.

\subsubsection*{Correspondence with the End Matter table and scope of the results}
\label{sec:micro-summary}

Substituting the six integrand parities into Eq.~\eqref{eq:collinear-current-kernel-map} gives Table~\ref{tab:micro-three-symmetries}.
To facilitate comparison, we use the same column order, $\Uop,\mathcal A,\Theta$, and mechanism order as in the End Matter table.
Each group lists the parities of the full charge- and spin-current kernels, followed by the corresponding allowed or forbidden channels.

\begin{table}[!htbp]
  \centering
  \caption{Six clean dc mechanisms in the collinear, spin-conserving limit, corresponding to the End Matter table.
    Each group lists $(\pi_c,\pi_{s_z})$, the charge channel, and the spin channel.
    $+$ and $-$ denote $+1$ and $-1$; $\times$ and $\checkmark$ denote symmetry-forbidden and symmetry-allowed contributions, respectively.
    The three symmetries are imposed separately. Powers of $\tau_r$ give the leading clean dc scaling.}
  \label{tab:micro-three-symmetries}
  \small
  \renewcommand{\arraystretch}{1.22}
  \setlength{\tabcolsep}{3pt}
  \begin{tabular}{@{}lcc*{3}{ccc}@{}}
    \toprule
    & & & \multicolumn{3}{c}{$\Uop=[C_2^\perp\Vert P]$}
      & \multicolumn{3}{c}{$\mathcal A=\mathcal T\tau$}
      & \multicolumn{3}{c}{$\Theta=[C_2^\perp\mathcal T\Vert E]$}\\
    \cmidrule(lr){4-6}\cmidrule(lr){7-9}\cmidrule(l){10-12}
    Mechanism & $N$ & Scaling
      & $(\pi_c,\pi_{s_z})$ & Charge & Spin
      & $(\pi_c,\pi_{s_z})$ & Charge & Spin
      & $(\pi_c,\pi_{s_z})$ & Charge & Spin\\
    \midrule
    Drude & 1 & $\tau_r$
      & $(+,-)$ & $\checkmark$ & $\usym{2613}$
      & $(+,-)$ & $\checkmark$ & $\usym{2613}$
      & $(+,+)$ & $\checkmark$ & $\checkmark$\\
    BC & 1 & $\tau_r^0$
      & $(+,-)$ & $\checkmark$ & $\usym{2613}$
      & $(-,+)$ & $\usym{2613}$ & $\checkmark$
      & $(-,-)$ & $\usym{2613}$ & $\usym{2613}$\\
    ND & 2 & $\tau_r^2$
      & $(-,+)$ & $\usym{2613}$ & $\checkmark$
      & $(-,+)$ & $\usym{2613}$ & $\checkmark$
      & $(-,-)$ & $\usym{2613}$ & $\usym{2613}$\\
    BCD & 2 & $\tau_r$
      & $(-,+)$ & $\usym{2613}$ & $\checkmark$
      & $(+,-)$ & $\checkmark$ & $\usym{2613}$
      & $(+,+)$ & $\checkmark$ & $\checkmark$\\
    inter-QMD & 2 & $\tau_r^0$
      & $(-,+)$ & $\usym{2613}$ & $\checkmark$
      & $(-,+)$ & $\usym{2613}$ & $\checkmark$
      & $(-,-)$ & $\usym{2613}$ & $\usym{2613}$\\
    intra-QMD & 2 & $\tau_r^0$
      & $(-,+)$ & $\usym{2613}$ & $\checkmark$
      & $(-,+)$ & $\usym{2613}$ & $\checkmark$
      & $(-,-)$ & $\usym{2613}$ & $\usym{2613}$\\
    \bottomrule
  \end{tabular}
\end{table}

Under $\Uop$, both linear mechanisms satisfy $\chi_\uparrow^X=\chi_\downarrow^X$, while all four second-order mechanisms satisfy $\chi_\uparrow^X=-\chi_\downarrow^X$.
The six mechanisms therefore realize the odd-order charge and even-order spin selection rules of Sec.~\ref{sec:covariance}.

Under $\mathcal A$, Drude, ND, and both QMD contributions obey the same rules as under $\mathcal U$, whereas antiunitary complex conjugation reverses the BC and BCD channel selection. Thus, $\mathcal A$ allows linear spin BC and second-order charge BCD responses despite sharing the same odd-parity bands with $\mathcal U$.

Under $\Theta$, BC, ND, and both QMD terms are already forbidden within each spin block.
For Drude and BCD, however, $\Theta$ alone does not determine the allowed charge and spin tensor components.
In altermagnets with a spin-group operation $O_s=[C_2^\perp\Vert O]$ relating opposite spins, the $N$th-order response satisfies
$\chi_\downarrow^{X,(N)}=\mathscr R_O^{(N+1)}\chi_\uparrow^{X,(N)}$.
Here $\mathscr R_O^{(N+1)}$ applies the orthogonal matrix of $O$ to the current index and all electric-field indices. Thus
\begin{equation}
  \begin{aligned}
    \chi_c^{X,(N)}
      &=(1+\mathscr R_O^{(N+1)})\chi_\uparrow^{X,(N)},\\
    \chi_{s_z}^{X,(N)}
      &=(1-\mathscr R_O^{(N+1)})\chi_\uparrow^{X,(N)}.
  \end{aligned}
  \label{eq:micro-Os-channels}
\end{equation}
The full magnetic point group is needed to determine which components survive.
The tensor structure of an individual mechanism can impose additional zeros:
for example, $\Omega_{ii}=0$ eliminates the fully longitudinal collinear BC and BCD components,
and Eq.~\eqref{eq:micro-inter-QMD} vanishes for $i=j=k$ because $2-1-1=0$.

Finally, the constraints that an antiunitary operation imposes on equilibrium quantities cannot replace unitary dynamical covariance.
For $\mathcal A$, for example, a uniform electric field transforms as $\bE^{\rm rev}(t)=\bE(-t)$.
The translation adds only a scalar to the length-gauge Hamiltonian and thus leaves commutators unchanged;
it does not produce the field $-\bE(t)$ required by $\Uop$.
Antiunitarity also maps retarded propagators to advanced propagators:
\begin{equation}
  \hat G^{R/A}(\epsilon,\bk)
    =[\epsilon-\hat H_0(\bk)\pm i\gamma]^{-1},\qquad
  \hat{\mathcal A} \hat G^R(\epsilon,\bk)\hat{\mathcal A}^{-1}
    =\hat G^A(\epsilon,-\bk),\qquad \gamma>0.
  \label{eq:micro-antiunitary-causality}
\end{equation}
$\Theta$ likewise reverses the causal direction.
The dc constraints for these antiunitary symmetries follow from the mechanism formulas analyzed here, while the all-order and harmonic separation theorems rely on the unitary dynamical covariance established in Sec.~\ref{sec:covariance}.

\section{Memory integrals along characteristics and nonperturbative periodic currents}
\label{sec:characteristic-memory}

\subsection{Intraband relaxation dynamics and the periodic steady state}
\label{sec:memory-assumptions}

The full periodic response in the main text is evaluated within the intraband relaxation-time approximation (RTA).
Let $\nu$ label all bands, including the spin degrees of freedom, and let $f_\nu(\bk,t)$ denote their occupations.
We retain electric-field-induced intraband acceleration and occupation relaxation, neglect interband transitions and coherence, and use the same constant relaxation time $\tau_r$ for all bands.
For a spatially uniform electric field $\bE(t)=E(t)\hat{\bm x}$, the distribution is independent of position.
Using the semiclassical acceleration theorem $\hbar\dot k_x=-eE(t)$, the Boltzmann equation reduces to
\begin{equation}
  \left[\partial_t-\frac{eE(t)}{\hbar}\partial_{k_x}\right]f_\nu(\bk,t)
  =-\frac{f_\nu(\bk,t)-f_\nu^{(0)}(\bk)}{\tau_r},
  \qquad
  f_\nu^{(0)}(\bk)=\frac{1}{e^{\beta[\varepsilon_\nu(\bk)-\mu]}+1}.
  \label{eq:memory-rta}
\end{equation}
Here the electron charge is $-e$ ($e>0$), $\mu$ and $\beta$ are the chemical potential and inverse temperature, respectively, $\tau_r>0$, and $f_\nu^{(0)}$ is the time-independent reference Fermi--Dirac distribution.
At degeneracies, the occupations must be assigned in a symmetry-compatible manner within the degenerate subspace.
We solve Eq.~\eqref{eq:memory-rta} by the method of characteristics~\cite{maccallumKineticIntegral1963,peresOpticalBistability2014}.

For a fixed observation time $t$ and momentum $\bk$, parameterize the trajectory $\bm K(l)$ ending at $(\bk,t)$ by $l\leq t$.
Its terminal condition and equation of motion are
\begin{equation}
  \bm K(t)=\bk,\qquad
  \frac{d\bm K(l)}{dl}=-\frac{eE(l)}{\hbar}\hat{\bm x},\qquad
  \bm K(l)=\bk+\frac{e}{\hbar}\hat{\bm x}\int_l^t dl'\,E(l').
  \label{eq:memory-characteristic}
\end{equation}
The transverse momentum remains constant along the trajectory.
Defining the occupation along the trajectory as $F_\nu(l)=f_\nu[\bm K(l),l]$, the chain rule gives
\begin{equation}
  \begin{aligned}
    \frac{dF_\nu(l)}{dl}
      &=\left[\partial_l f_\nu(\bk,l)
        +\frac{d\bm K(l)}{dl}\cdot\nabla_{\bk}f_\nu(\bk,l)
        \right]_{\bk=\bm K(l)}\\
      &=-\frac{F_\nu(l)-f_\nu^{(0)}[\bm K(l)]}{\tau_r}.
  \end{aligned}
  \label{eq:memory-characteristic-ode}
\end{equation}
Equation~\eqref{eq:memory-rta} thus reduces to a first-order linear ordinary differential equation along each trajectory.
Although $f_\nu^{(0)}(\bk)$ has no explicit time dependence, its value along the trajectory, $f_\nu^{(0)}[\bm K(l)]$, varies with $l$.

Moving $F_\nu/\tau_r$ to the left-hand side and multiplying by the integrating factor $e^{l/\tau_r}$ yields
\begin{equation}
  \frac{d}{dl}\left[e^{l/\tau_r}F_\nu(l)\right]
    =\frac{e^{l/\tau_r}}{\tau_r}f_\nu^{(0)}[\bm K(l)].
  \label{eq:memory-integrating-factor}
\end{equation}
Integrating first from a finite initial time $t_0$ to $t$, and using $F_\nu(t)=f_\nu(\bk,t)$, gives
\begin{equation}
  \begin{aligned}
    f_\nu(\bk,t)={}&e^{-(t-t_0)/\tau_r}f_\nu[\bm K(t_0),t_0]\\
      &+\int_{t_0}^{t}\frac{dl}{\tau_r}e^{-(t-l)/\tau_r}
        f_\nu^{(0)}\!\left[\bk+\frac{e}{\hbar}\hat{\bm x}
          \int_l^t dl'\,E(l')\right].
  \end{aligned}
  \label{eq:memory-finite-initial-time}
\end{equation}
The first term carries the initial occupation and decays on the timescale $\tau_r$; the second accumulates the relaxation source along the trajectory.
Because physical occupations are bounded, the first term vanishes as $t_0\to-\infty$, leaving the retarded solution~\cite{maccallumKineticIntegral1963}
\begin{equation}
  f_\nu(\bk,t)=\int_{-\infty}^{t}\frac{dl}{\tau_r}e^{-(t-l)/\tau_r}
    f_\nu^{(0)}\!\left[\bk+\frac{e}{\hbar}\hat{\bm x}
      \int_l^t dl'\,E(l')\right].
  \label{eq:memory-past-time-solution}
\end{equation}
Changing variables to the elapsed time $u=t-l\geq0$ gives the causal memory solution used in the main text:
\begin{equation}
  f_\nu(\bk,t)=\int_0^\infty\frac{du}{\tau_r}e^{-u/\tau_r}
    f_\nu^{(0)}\!\left[\bk+q(t,u)\hat{\bm x}\right],
  \qquad
  q(t,u)=\frac{e}{\hbar}\int_{t-u}^{t}dl\,E(l).
  \label{eq:memory-retarded-solution}
\end{equation}
The exponential kernel is normalized, $\int_0^\infty du\,e^{-u/\tau_r}/\tau_r=1$.
The occupation is therefore a normalized weighted average of the reference distribution along the past trajectory, with $\tau_r$ setting the effective memory time.
At zero field, $q=0$, and Eq.~\eqref{eq:memory-retarded-solution} reduces exactly to $f_\nu^{(0)}(\bk)$.
For a given electric field and reference distribution, the difference $\Delta f_\nu$ between any two solutions satisfies
\begin{equation}
  \Delta f_\nu(\bk,t)=e^{-(t-t_0)/\tau_r}
    \Delta f_\nu[\bm K(t_0),t_0].
  \label{eq:memory-transient-decay}
\end{equation}
If $E(t+T)=E(t)$, then $q(t+T,u)=q(t,u)$, so Eq.~\eqref{eq:memory-retarded-solution} is also $T$ periodic.
Together with the exponential decay in Eq.~\eqref{eq:memory-transient-decay}, this establishes the memory solution as the unique attracting periodic steady state.

For the monochromatic drive $E(t)=E_0\cos\Omega t$ used in the main text, direct integration yields
\begin{equation}
  q(t,u)=K_0[\sin\Omega t-\sin\Omega(t-u)],
  \qquad K_0=\frac{eE_0}{\hbar\Omega},\qquad T=\frac{2\pi}{\Omega}.
  \label{eq:memory-sinusoidal-shift}
\end{equation}
Substituting this displacement into Eq.~\eqref{eq:memory-retarded-solution} gives the steady-state occupation at any drive phase without assuming $\Omega\tau_r\ll1$, and thus retains the relaxation-induced phase lag at finite frequency.
In lattice units, $k_1=ak_x$, the corresponding displacement is $a q$, and the dimensionless field amplitude in the main text is $\widetilde E_0=aK_0$.

The current follows by inserting the occupation into the band sum.
Let $\zeta_{i\nu}^{\alpha}(\bk)$ denote a current kernel with velocity units, where $\alpha\in\{c,s_z\}$.
The charge kernel is $\zeta_{i\nu}^{c}=v_{i\nu}$; the spin kernel is specified in Sec.~\ref{sec:memory-current}.
Consistent with our definition of field-induced currents, we subtract the zero-field background and write
\begin{equation}
  J_{\alpha,i}(t)=-e\sum_\nu\intk\zeta_{i\nu}^{\alpha}(\bk)
    [f_\nu(\bk,t)-f_\nu^{(0)}(\bk)].
  \label{eq:memory-current-from-occupation}
\end{equation}
Here $d$ is the spatial dimension, and the integral covers the full Brillouin zone.
Substituting Eq.~\eqref{eq:memory-retarded-solution} and using the normalization of the memory kernel gives
\begin{equation}
  \begin{aligned}
    J_{\alpha,i}(t)={}&-e\int_0^\infty\frac{du}{\tau_r}e^{-u/\tau_r}
      \sum_\nu\intk\zeta_{i\nu}^{\alpha}(\bk)\\
      &\times\left\{f_\nu^{(0)}[\bk+q(t,u)\hat{\bm x}]-f_\nu^{(0)}(\bk)\right\}.
  \end{aligned}
  \label{eq:memory-current-shifted-occupation}
\end{equation}

An equivalent form follows by changing variables to $\bk'=\bk+q(t,u)\hat{\bm x}$ in the first term of Eq.~\eqref{eq:memory-current-shifted-occupation} at each fixed $u$.
Brillouin-zone periodicity allows the shifted integration domain to be mapped back to the original cell.
Relabeling $\bk'$ as $\bk$ then gives
\begin{equation}
  \begin{aligned}
    J_{\alpha,i}(t)={}&-e\sum_\nu\intk f_\nu^{(0)}(\bk)
      \int_0^\infty\frac{du}{\tau_r}e^{-u/\tau_r}\\
      &\times\left\{\zeta_{i\nu}^{\alpha}[\bk-q(t,u)\hat{\bm x}]
        -\zeta_{i\nu}^{\alpha}(\bk)\right\}.
  \end{aligned}
  \label{eq:memory-current-shifted-kernel}
\end{equation}
The two forms place the momentum displacement in the occupation or in the current kernel, respectively.
In the charge channel, the latter is a memory average of the shifted band velocity.
Retaining the full displacement $q(t,u)\propto E_0$ includes all orders in the electric-field amplitude within the intraband RTA.

\subsection{Currents in collinear and noncollinear systems}
\label{sec:memory-current}

We evaluate the charge and $z$-polarized spin currents from the same band occupations, using the following current kernels with velocity units:
\begin{equation}
  \begin{aligned}
    \zeta_{i\nu}^{c}(\bk)
      &=\langle u_{\nu\bk}|\hat v_i|u_{\nu\bk}\rangle
       =\frac{1}{\hbar}\partial_{k_i}\varepsilon_\nu(\bk),\\
    \zeta_{i\nu}^{s_z}(\bk)
      &=\frac{1}{\hbar}\langle u_{\nu\bk}|
        \{\hat v_i,\hat s_z\}|u_{\nu\bk}\rangle.
  \end{aligned}
  \label{eq:memory-current-kernels}
\end{equation}
The full anticommutator matrix element must be retained in noncollinear systems.
When $\hat s_z$ is conserved, the spin kernel reduces to $+v_{i\nu}$ and $-v_{i\nu}$ in the spin-up and spin-down blocks, respectively, recovering $J_c=J_\uparrow+J_\downarrow$ and $J_{s_z}=J_\uparrow-J_\downarrow$.
As in the main text, the spin current is expressed in charge-equivalent units; multiplication by $-\hbar/(2e)$ gives the angular-momentum current.

Inserting the kernels of Eq.~\eqref{eq:memory-current-kernels} into Eq.~\eqref{eq:memory-current-shifted-occupation} or Eq.~\eqref{eq:memory-current-shifted-kernel} gives the field-induced currents for both collinear and noncollinear systems from the same occupation solution.
For Floquet effective bands, we additionally assume quasiequilibrium Fermi--Dirac reference occupations with common $\mu$ and $\beta$.

\subsection{Half-period symmetry and harmonic selection}
\label{sec:memory-half-period}

Harmonics are defined using the positive-exponent convention in Eq.~\eqref{eq:full-current-fourier}.
The $n=0$ component is the field-induced dc rectification, and $|J_{\alpha,i}^{n\Omega}|$ is the modulus of the complex Fourier coefficient.
Because the momentum displacement is retained in full, a given output harmonic can contain contributions from multiple response orders of the same parity.

When $\Uop$ is preserved, the charge kernel is odd and the spin kernel is even under $\Uop$.
A drive that reverses sign after half a period satisfies $q(t+T/2,u)=-q(t,u)$.
At fixed elapsed time $u$, changing variables to the symmetry-related bands at $-\mathbf k$ shows that the charge contribution is odd in $q$ and the spin contribution is even in $q$. The memory integral preserves these parities.
This argument requires neither conserved $\hat s_z$ nor $\Omega\tau_r\ll1$.

\section{Numerical implementation and computational details}
\label{sec:numerical-details}

This section describes the calculations without SOC used in the main text; calculations including SOC are presented in Sec.~\ref{sec:soc}.

\subsection{\texorpdfstring{First-principles calculations, Wannier models, and Floquet engineering}{First-principles calculations, Wannier models, and Floquet engineering}}
\label{sec:numerical-dft}

First-principles calculations were performed using VASP~\cite{kresseEfficientIterative1996}.
The electron--ion interaction was treated with the projector augmented-wave (PAW) method~\cite{blochlProjectorAugmented1994,kresseUltrasoftPAW1999}.
Exchange and correlation were described by the Perdew--Burke--Ernzerhof (PBE) functional~\cite{perdewGeneralizedGradient1996}.
For both materials, static self-consistent calculations were carried out at fixed atomic structures without spin-orbit coupling, using Gaussian smearing of $0.05~\mathrm{eV}$. Monolayer MnPS$_3$ exhibits the collinear N\'eel antiferromagnetic order, with $U_{\mathrm{eff}}=5.0~\mathrm{eV}$ applied to the Mn $3d$ orbitals in the Dudarev formulation~\cite{dudarevLSDAU1998}. The plane-wave energy cutoff, self-consistent energy-convergence threshold, and $\Gamma$-centered $k$ mesh were $500~\mathrm{eV}$, $10^{-10}~\mathrm{eV}$, and $12\times12\times1$, respectively. Monolayer VCl$_2$ exhibits coplanar $120^\circ$ antiferromagnetic order within a $\sqrt3\times\sqrt3$ magnetic cell, with $U=1.8~\mathrm{eV}$ and $J=0.8~\mathrm{eV}$ applied to the V $3d$ orbitals in the Liechtenstein formulation~\cite{liechtensteinDensityFunctional1995}. The corresponding parameters were $600~\mathrm{eV}$, $10^{-8}~\mathrm{eV}$, and $14\times14\times1$.

Tight-binding models were constructed using orbital projections in Wannier90~\cite{pizziWannier902020}. For MnPS$_3$, Mn $d$, S $p$, and P $p$ projectors yield $34$ orbitals in each conserved spin block. The spinor basis of VCl$_2$ contains $66$ orbitals obtained from V $d$ and Cl $p$ projectors.

Following the effective-Hamiltonian treatment of light-driven monolayer MnPS$_3$ in Ref.~\cite{huangLightInducedOddParity2026}, we construct the pump-dressed bands using the van Vleck expansion through first order in $1/\omega_p$. The Peierls-coupled Hamiltonian retains Fourier components $|m|\leq1$, with $\hbar\omega_p=10~\mathrm{eV}$ and $eA_p/\hbar=0.30~\text{\AA}^{-1}$. The full Wannier velocity and spin-current operators are dressed consistently to the same order in $1/\omega_p$.

\subsection{Discretization of the memory integrals along characteristics}
\label{sec:numerical-memory}

For the minimal model and VCl$_2$, the substitution $x=u/\tau_r$ in Eq.~\eqref{eq:memory-current-shifted-occupation} gives a semi-infinite memory integral with weight $e^{-x}$, which was evaluated using Gauss--Laguerre (GL) quadrature. For MnPS$_3$, the memory integral was folded into one probe period and evaluated using Gauss--Legendre quadrature. The Brillouin zone was sampled on a uniform two-dimensional mesh. For MnPS$_3$ and VCl$_2$, the displacement-dependent current kernels were tabulated at $2049$ uniformly spaced nodes over $[-2K_0,2K_0]$ and evaluated using cubic-spline interpolation.
Currents were sampled uniformly over one probe period, and the Fourier coefficients were obtained by discretizing Eq.~\eqref{eq:full-current-fourier}. All plotted harmonic amplitudes are moduli of the complex Fourier coefficients. The principal parameters are listed in Table~\ref{tab:numerical-memory}, where $N_t$ is the number of time points per period and $N_{\mathrm{mem}}$ counts the quadrature nodes for the memory integral.

\begin{table}[!htbp]
  \centering
  \caption{Numerical parameters used for the periodic currents in the main text. Each $k$ mesh covers the full two-dimensional Brillouin zone.}
  \label{tab:numerical-memory}
  \small
  \renewcommand{\arraystretch}{1.18}
  \setlength{\tabcolsep}{7pt}
  \begin{tabular}{@{}lcccc@{}}
    \toprule
    System and main-text figure & $k$ mesh & $N_t$ & $N_{\mathrm{mem}}$ & $\Omega\tau_r$\\
    \midrule
    Minimal model, Fig.~2(c,d) & $101\times101$ & 512 & 96 & 1\\
    Minimal model, Fig.~2(e,f) & $61\times61$ & 256 & 80 & 1\\
    Monolayer MnPS$_3$, Fig.~3(c,d) & $140\times140$ & 4096 & 2048 & 1.25\\
    Monolayer VCl$_2$, Fig.~3(e)--(h) & $336\times336$ & 1024 & 256 & 1\\
    \bottomrule
  \end{tabular}
\end{table}
\FloatBarrier

For the minimal model, the reference field amplitude is $\widetilde E_0=1.2$, and the field-amplitude scan covers $0.02$--$1.5$. For MnPS$_3$, the probe is polarized along $x$, with $\tau_r=10~\mathrm{fs}$, $K_0=1.8~\text{\AA}^{-1}$, and temperature $T_{e}=10~\mathrm{K}$. For VCl$_2$, $K_0=0.0118~\text{\AA}^{-1}$, $\tau_r=10~\mathrm{fs}$, and temperature $T_{e}=10~\mathrm{K}$.

\subsection{Symmetrization of the two material models}
\label{sec:numerical-symmetrization}

For MnPS$_3$, the Hamiltonian, in-plane position connection, and conserved spin operator were jointly averaged under $\Uop=[C_{2x}\Vert P]$ in the Wannier representation, with consistent treatment of orbital inversion phases, lattice translations, and Wannier centers. The full velocity and anticommutator spin-current matrices were then reconstructed from these symmetrized operators. The results in Fig.~3(a)--(d) of the main text use this symmetrized input. A $384\times384$ mesh covering the full Brillouin zone, supplemented by a local search, gives a maximum energy change of approximately $2.09\times10^{-7}~\mathrm{eV}$ ($0.209~\mu\mathrm{eV}$) between the bare bands before and after symmetrization.

For VCl$_2$, the Hamiltonian, position connection, and all three spin components were jointly averaged over the group generated by $[C_{2x}\Vert P]$ and $[E\Vert C_{2y}]$ in the same Wannier basis.
The transformation of the position connection retains the momentum-derivative term of the representation matrix.
The velocity and full anticommutator spin-current matrices were then recalculated consistently.
The maximum band-energy correction was approximately $0.016~\mathrm{meV}$, and the relative changes in all allowed complex harmonic coefficients for orders $0$--$4$ were below $0.04\%$.

\section{Effects of spin-orbit coupling}
\label{sec:soc}

Spin-orbit coupling (SOC) affects charge--spin harmonic separation through its action on the protecting symmetry $\Uop$.
We first compare symmetry-preserving and symmetry-breaking SOC in the four-band model, then examine the robustness of harmonic separation in monolayer VCl$_2$ and MnPS$_3$.
Throughout this section, we apply $E_x(t)=E_0\cos\Omega t$ and evaluate the periodic intraband response using the RTA of Sec.~\ref{sec:characteristic-memory}, with a common relaxation time $\tau_r$ for all bands.
We compute the currents from the full operators in Eq.~\eqref{eq:current-definitions}, subtract the equilibrium background, and use the Fourier convention of Eq.~\eqref{eq:full-current-fourier}.

\subsection{Symmetry-preserving and symmetry-breaking SOC in the minimal odd-parity model}
\label{sec:soc-model}

We extend the four-band $p$-wave Hamiltonian of the main text by adding two spin-dependent hopping terms,
\begin{equation}
  H(\bk;\lambda_+,\lambda_-)
    =H_0(\bk)+\lambda_+\sin k_1\,\tau_0\sigma_y
                 +\lambda_-\sin k_1\sin^2\!\frac{k_2}{2}\,\tau_z\sigma_z,
  \label{eq:soc-H}
\end{equation}
where $H_0(\bk)\equiv H_\delta(\bk)|_{\delta=0}$, $(k_1,k_2)=a(k_x,k_y)$, and $a$ is the lattice constant.
The Pauli matrices $\tau_i$ and $\sigma_i$ act in sublattice and spin space, respectively.
The $\lambda_+$ term describes sublattice-uniform hopping along $x$, and the $\lambda_-$ term describes sublattice-staggered hopping along $x$ and the two diagonals.
In the basis $(A\uparrow,A\downarrow,B\uparrow,B\downarrow)$, $U_{\Uop}=\tau_x\sigma_x$ gives
\begin{equation}
  U_{\Uop}H(\bk;\lambda_+,\lambda_-)U_{\Uop}^{\dagger}
    =H(-\bk;\lambda_+,-\lambda_-).
  \label{eq:soc-covariance}
\end{equation}
Thus, the $\lambda_+$ term preserves $\Uop$ while mixing the two $z$-spin components, whereas the $\lambda_-$ term breaks $\Uop$ and conserves $\hat s_z$ when $\lambda_+=0$.
These two terms separate the effects of spin mixing and symmetry breaking.

\begin{figure}[!htbp]
  \centering
  \includegraphics[width=0.9\linewidth,clip]{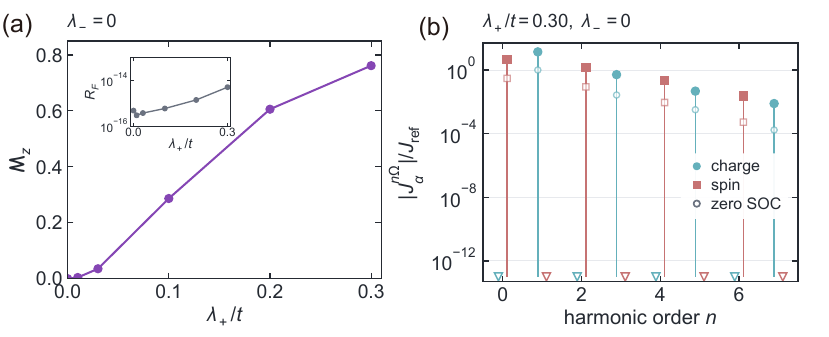}
  \caption{Charge--spin harmonic separation with symmetry-preserving SOC ($\lambda_-=0$).
    (a) Thermally weighted spin-mixing measure $\mathcal M_z$ versus $\lambda_+/t$.
    Inset: normalized forbidden residual
    $R_F=\max_{m=0,\ldots,3}\{|J_c^{2m\Omega}|,|J_{s_z}^{(2m+1)\Omega}|\}/J_{\mathrm{ref}}$.
    (b) Harmonic spectrum at $\lambda_+/t=0.30$.
    Cyan circles and red squares denote charge and spin amplitudes; pale open markers show the allowed harmonics at zero SOC.
    Open inverted triangles mark components below $10^{-13}J_{\mathrm{ref}}$.
    Parameters are $t=\Delta_z=1~\mathrm{eV}$, $t_0=\lambda_p=0.3~\mathrm{eV}$, $\mu=-3~\mathrm{eV}$,
    $k_BT_{\mathrm e}=0.025852~\mathrm{eV}$, $\delta=0$, $\Omega\tau_r=1$, and
    $\widetilde E_0=aeE_0/(\hbar\Omega)=1.2$.
    Both channels are normalized by $J_{\mathrm{ref}}=|J_c^\Omega(0,0)|$.}
  \label{fig:soc-plus}
\end{figure}

We calculate the longitudinal currents using $\hat v_x=(a/\hbar)\partial_{k_1}H$ and $\hat s_z=(\hbar/2)\tau_0\sigma_z$, and suppress the spatial index $x$ below.
Both SOC scans use the same zero-SOC normalization, $J_{\mathrm{ref}}=|J_c^\Omega(0,0)|$.
The calculation uses a $201\times201$ momentum mesh, $1024$ time points per period, and $256$ Gauss--Legendre nodes for the memory integral folded into one drive period.
Model and drive parameters are given in Fig.~\ref{fig:soc-plus}.

For $\lambda_-=0$, we quantify spin mixing near the Fermi level by
\begin{equation}
  \mathcal M_z=
  \frac{\displaystyle\sum_{\nu,\bk}f_{\nu\bk}(1-f_{\nu\bk})
    \left[1-\langle\Sigma_z\rangle_{\nu\bk}^{\,2}\right]}
       {\displaystyle\sum_{\nu,\bk}f_{\nu\bk}(1-f_{\nu\bk})},
  \qquad \Sigma_z=\frac{2\hat s_z}{\hbar},
  \label{eq:soc-mixing}
\end{equation}
where $\nu$ runs over the four bands and $f_{\nu\bk}$ is the equilibrium Fermi occupation.
Figure~\ref{fig:soc-plus}(a) shows that $\mathcal M_z$ reaches approximately $0.762$ at $\lambda_+/t=0.30$, indicating substantial spin mixing.
The preserved $\Uop$ symmetry and the half-period antisymmetry of the drive enforce the harmonic selection rule in Eq.~\eqref{eq:complete-harmonic-selection}.
The largest forbidden residual over the scan is approximately $5.2\times10^{-15}J_{\mathrm{ref}}$.
At $\lambda_+/t=0.30$, the spectrum contains only odd charge harmonics and even spin harmonics, including dc spin rectification [Fig.~\ref{fig:soc-plus}(b)].
The allowed amplitudes vary substantially with SOC.

\begin{figure}[!htbp]
  \centering
  \includegraphics[width=0.90\linewidth,clip]{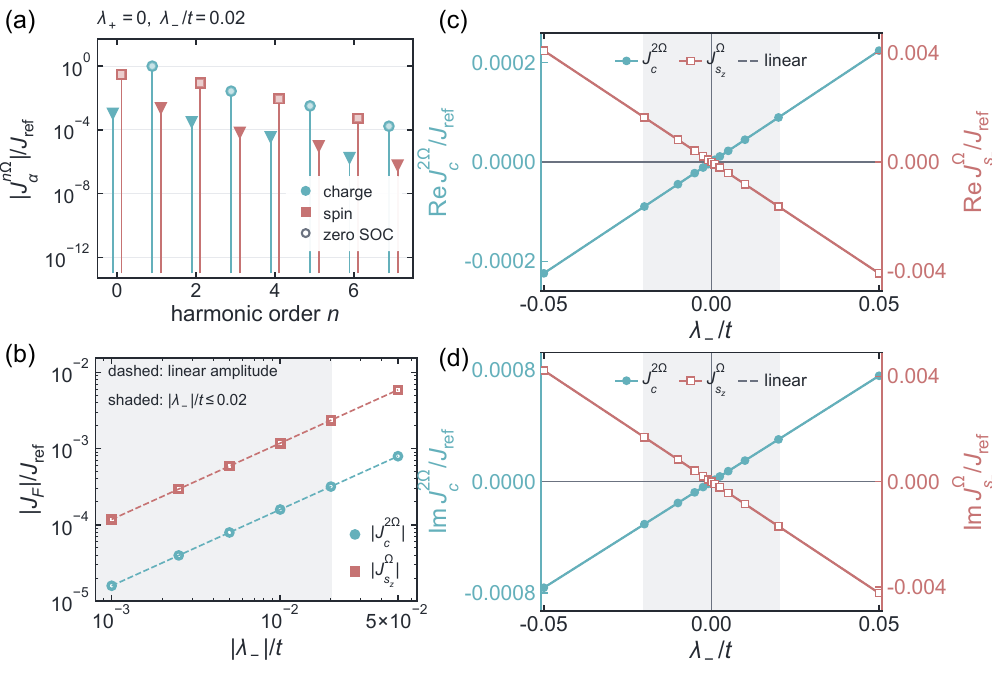}
  \caption{Forbidden harmonics induced by the $\lambda_-$ term in Eq.~\eqref{eq:soc-H} at $\lambda_+=0$.
    Model and drive parameters and $J_{\rm ref}$ are as in Fig.~\ref{fig:soc-plus}.
    (a) Spectrum for $n=0,\ldots,7$ at $\lambda_-/t=0.02$.
    Cyan and red denote charge and spin. Filled inverted triangles mark activated harmonics, and pale open circles and squares show the allowed zero-SOC harmonics.
    (b) $|J_F|/J_{\rm ref}$ for $J_F=J_c^{2\Omega},J_{s_z}^{\Omega}$ versus $|\lambda_-|/t$, with large filled and small open markers for positive and negative $\lambda_-$, respectively.
    (c,d) Real and imaginary parts of $J_F/J_{\rm ref}$, with charge on the left axis and spin on the right.
    Dashed lines in (b)--(d) show the corresponding modulus or component of $C_1\lambda_-/J_{\rm ref}$. Shading marks $|\lambda_-|/t\leq0.02$.}
  \label{fig:soc-minus}
\end{figure}

For $\lambda_+=0$, the symmetry-breaking term activates even charge and odd spin harmonics [Fig.~\ref{fig:soc-minus}(a)].
The longitudinal drive shifts only $k_1$, so the quadratic node at $k_2=0$ keeps the perturbation weak for the thermally occupied states near that line throughout the drive.
At $\lambda_-/t=0.02$, the representative amplitudes are
$|J_c^{2\Omega}|/J_{\mathrm{ref}}\simeq3.17\times10^{-4}$ and
$|J_{s_z}^{\Omega}|/J_{\mathrm{ref}}\simeq2.35\times10^{-3}$.
The activated spin fundamental reaches $2.72\%$ of the spin second harmonic.
Equation~\eqref{eq:soc-covariance} and the half-period shift of the drive give
\begin{equation}
  J_c^{n\Omega}(\lambda_-)=(-1)^{n+1}J_c^{n\Omega}(-\lambda_-),
  \qquad
  J_{s_z}^{n\Omega}(\lambda_-)=(-1)^nJ_{s_z}^{n\Omega}(-\lambda_-).
  \label{eq:soc-parameter-parity}
\end{equation}
The formerly forbidden harmonics are therefore odd functions of $\lambda_-$.
For $J_F\in\{J_c^{2\Omega},J_{s_z}^{\Omega}\}$, the small-SOC expansion is
\begin{equation}
  J_F(\lambda_-)=C_1\lambda_-+C_3\lambda_-^3+O(\lambda_-^5),
  \label{eq:soc-onset}
\end{equation}
with complex coefficients $C_1$ and $C_3$.
At $\widetilde E_0=1.2$, Fig.~\ref{fig:soc-minus}(b) shows the linear onset, and panels (c) and (d) show the sign reversals of the real and imaginary parts, respectively.
We fit $C_1$ and $C_3$ to the complex amplitudes for $0<|\lambda_-|/t\leq0.005$.
For the points with $0<|\lambda_-|/t\leq0.02$, the relative deviation $|J_F-C_1\lambda_-|/|C_1\lambda_-|$ remains below $0.007\%$ in both channels. The activated harmonics therefore provide a probe of weak SOC-induced breaking of $\Uop$.

\FloatBarrier

\subsection{Harmonic separation in monolayer \texorpdfstring{VCl$_2$}{VCl2} with SOC}
\label{sec:soc-vcl2}

Monolayer VCl$_2$ retains a strong separation of charge and spin harmonics when its intrinsic SOC is included.
Starting from the in-plane $120^\circ$ noncollinear antiferromagnetic configuration of the main text, we perform a self-consistent SOC calculation at fixed atomic structure in VASP and construct a $66$-dimensional spinor Wannier model from V $d$ and Cl $p$ orbitals.
The resulting Hamiltonian and spin matrices are used directly in the transport calculation.
For the $x$-directed drive, we evaluate the longitudinal charge current $J_{c,x}$ and transverse $z$-polarized spin current $J_{s_z,y}$.

\begin{figure}[!htbp]
  \centering
  \includegraphics[width=0.96\linewidth,clip]{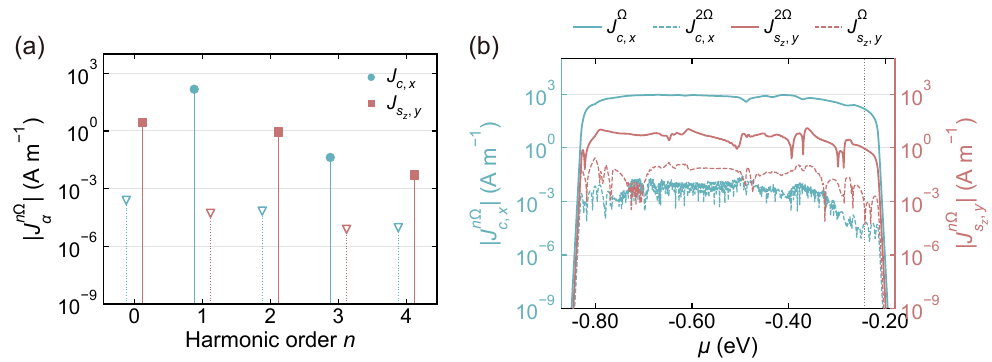}
  \caption{Charge--spin harmonic separation in monolayer VCl$_2$ with SOC.
    (a) Harmonic amplitudes of $J_{c,x}$ and $J_{s_z,y}$ at $\mu-E_F=-0.242~\mathrm{eV}$.
    Cyan circles and red squares denote charge and spin components allowed by the zero-SOC selection rule; open inverted triangles denote the formerly forbidden components.
    (b) Fundamental charge and second-harmonic spin amplitudes, together with their formerly forbidden counterparts, versus chemical potential.
    Solid (dashed) curves denote allowed (formerly forbidden) components; the vertical dotted line marks the filling in (a).
    The energy zero is $E_F$ of the self-consistent SOC calculation.
    Parameters are a $140\times140$ transport $k$ mesh, $T_{\mathrm e}=10~\mathrm K$,
    $K_0=eE_0/(\hbar\Omega)=0.01183~\text{\AA}^{-1}$, and $\Omega\tau_r=1$.
    Amplitudes are moduli of the field-induced complex Fourier coefficients, with spin currents in charge-equivalent units.}
  \label{fig:vcl2-soc}
\end{figure}

At the representative chemical potential $\mu-E_F=-0.242~\mathrm{eV}$ used in the main text, odd charge harmonics ($n=1,3$) and even spin harmonics ($n=0,2,4$) dominate the spectrum [Fig.~\ref{fig:vcl2-soc}(a)].
The transverse dc spin rectification and second spin harmonic remain sizable, while components forbidden in the absence of SOC are strongly suppressed.
We quantify this separation over $n=0,\ldots,4$ by
\begin{equation}
  R_c=
  \frac{\displaystyle\max_{n\in\{0,2,4\}}|J_{c,x}^{n\Omega}|}
       {\displaystyle\max_{n\in\{1,3\}}|J_{c,x}^{n\Omega}|},
  \qquad
  R_{s_z}=
  \frac{\displaystyle\max_{n\in\{1,3\}}|J_{s_z,y}^{n\Omega}|}
       {\displaystyle\max_{n\in\{0,2,4\}}|J_{s_z,y}^{n\Omega}|}.
  \label{eq:vcl2-soc-ratios}
\end{equation}
At this filling, $R_c=1.65\times10^{-6}$ and $R_{s_z}=1.88\times10^{-5}$, demonstrating that harmonic separation survives intrinsic SOC to high accuracy.

The chemical-potential scan in Fig.~\ref{fig:vcl2-soc}(b) shows that carrier filling tunes the dominant responses while the formerly forbidden components remain suppressed.
All $61$ sampled points in $-0.28\leq\mu-E_F\leq-0.22~\mathrm{eV}$ satisfy
$R_c<3.7\times10^{-6}$ and $R_{s_z}<2.6\times10^{-3}$.
This filling window supports harmonic-resolved detection of dc and even-harmonic spin currents together with odd-harmonic charge currents in a material with intrinsic SOC.

\subsection{Harmonic separation in monolayer MnPS$_3$ with SOC}
\label{sec:soc-mnps3}

For MnPS$_3$, we include SOC self-consistently in VASP.
Starting from the N\'eel state with moments along $z$, we construct a $68$-dimensional spinor Wannier model and use its unsymmetrized Hamiltonian and operators directly in the transport calculation.
The pump-dressed Hamiltonian and full current operators are evaluated to first order in $1/\omega_p$, retaining $m=0,\pm1$.

\begin{figure}[!htbp]
  \centering
  \includegraphics[width=0.96\linewidth,clip]{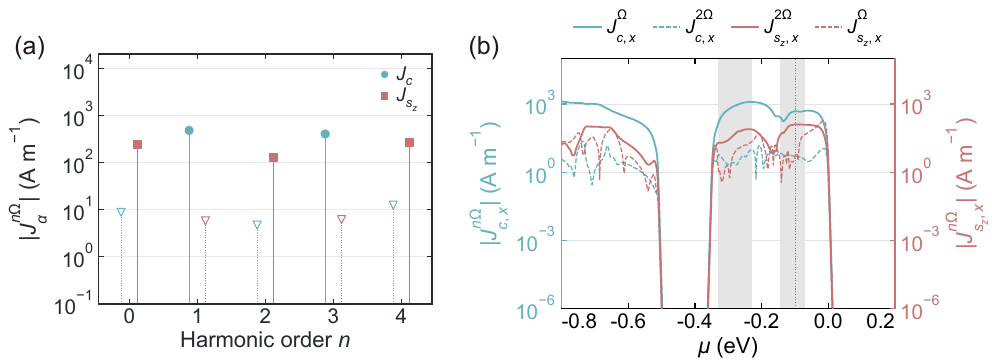}
  \caption{Charge--spin harmonic separation in Floquet monolayer MnPS$_3$ with SOC.
    (a) Harmonic amplitudes of $J_{c,x}$ and $J_{s_z,x}$ at $\mu\simeq-0.0986~\mathrm{eV}$.
    (b) Fundamental and second harmonics versus chemical potential, measured from the Floquet valence-band maximum.
    The vertical dotted line marks the filling in (a), and gray bands highlight representative filling intervals.
    Parameters are a $140\times140$ momentum mesh, $T_{\mathrm e}=10~\mathrm K$, $\tau_r=10~\mathrm{fs}$, $K_0=1.8~\text{\AA}^{-1}$, $\Omega\tau_r=1.25$, $\hbar\omega_p=10~\mathrm{eV}$, and $eA_p/\hbar=0.30~\text{\AA}^{-1}$.}
  \label{fig:mnps3-soc}
\end{figure}

At $\mu\simeq-0.0986~\mathrm{eV}$ relative to the Floquet valence-band maximum, odd charge harmonics and even spin harmonics, including dc spin rectification, dominate the response [Fig.~\ref{fig:mnps3-soc}(a)]. Using Eq.~\eqref{eq:vcl2-soc-ratios} with $J_{s_z,y}$ replaced by $J_{s_z,x}$ gives $R_c\simeq2.6\%$ and $R_{s_z}\simeq2.3\%$. The complementary harmonic pattern remains clearly resolved in the presence of intrinsic SOC.

\FloatBarrier

With SOC included, the maximum energy mismatch between $\Uop$-related band partners over the full Brillouin zone is approximately $5~\mathrm{meV}$ in VCl$_2$ and $60~\mathrm{meV}$ in MnPS$_3$.
This contrast reveals substantially weaker SOC-induced breaking of $\Uop$ in the band structure of VCl$_2$.
A qualitative microscopic picture follows from the spin orientation and orbital character.
The atomic spin--orbit interaction takes the form
\begin{equation}
  \hat H_{\mathrm{SOC}}
  =\sum_i\xi_i\hat{\bm L}_i\cdot\hat{\bm S}_i,
  \label{eq:atomic-soc}
\end{equation}
where $i$ labels atomic sites, $\xi_i$ is the corresponding SOC strength, and angular momenta are expressed in units of $\hbar$.
For $\Uop=[C_{2x}\Vert P]$, the spin rotation preserves $S_x$ and reverses $S_y$ and $S_z$.
Inversion permutes atomic sites and leaves orbital angular momentum unchanged as an axial vector.
After relabeling the inversion-related sites, the symmetry-even and symmetry-odd parts of SOC are therefore
\begin{equation}
  \begin{aligned}
    \hat H_+
    &=\frac{\hat H_{\mathrm{SOC}}+\Uop\hat H_{\mathrm{SOC}}\Uop^{-1}}{2}
      =\sum_i\xi_i\hat L_{xi}\hat S_{xi},\\
    \hat H_-
    &=\frac{\hat H_{\mathrm{SOC}}-\Uop\hat H_{\mathrm{SOC}}\Uop^{-1}}{2}
      =\sum_i\xi_i\left(\hat L_{yi}\hat S_{yi}+\hat L_{zi}\hat S_{zi}\right).
  \end{aligned}
  \label{eq:atomic-soc-parity}
\end{equation}
They satisfy $\Uop\hat H_\pm\Uop^{-1}=\pm\hat H_\pm$. The action of $\hat H_-$ on the band states controls the SOC-induced energy mismatch between $\Uop$-related partners.

In a local exchange-field picture, the two spin branches are polarized parallel and antiparallel to the local moment~\cite{Kosugi2014}.
SOC couples states both within and between these branches.
When the relevant interbranch energy separations exceed the SOC matrix elements, energy corrections from virtual transitions between the branches are suppressed by the corresponding energy denominators.

For the $z$-collinear MnPS$_3$ configuration studied here, $\hat L_z\hat S_z$ preserves the local spin direction. It couples orbital states within the same spin branch and shifts their energies without admixing the exchange-separated opposite-spin branch. The relevant $L_z$ matrix elements thus provide a direct channel for unequal SOC corrections to $\Uop$-related partner energies.

For the in-plane moments of VCl$_2$, the $z$ direction is perpendicular to every local moment. In the local exchange-field picture, $\hat L_z\hat S_z$ is off-diagonal in the local spin branches. Its direct intrabranch matrix element vanishes, and virtual interbranch mixing is suppressed by exchange splitting. This weakens the direct contribution of this channel to symmetry breaking.

The $\hat L_y\hat S_y$ term can still act within the same spin branch through the component of $S_y$ parallel to the local moment.
If the $L_y$ matrix elements between the relevant low-energy orbitals are weak, the associated contribution to the partner energy mismatch is also suppressed.
Together with the exchange suppression of the $L_zS_z$ channel, this provides a qualitative explanation for the smaller maximum partner energy mismatch over the full Brillouin zone in VCl$_2$ ($5~\mathrm{meV}$ compared with $60~\mathrm{meV}$ in MnPS$_3$).

\bibliography{SMref}